\documentclass[aps,superscriptaddress,amsmath,amssymb,floatfix,twocolumn,showpacs,amsfonts,longbibliography,reprint]{revtex4-2}
\usepackage{times}
\usepackage[varg]{txfonts}
\usepackage{textcomp}
\usepackage{graphicx}
\usepackage{subfigure}
\usepackage{color}
\usepackage[colorlinks=true,citecolor=blue,urlcolor=blue,linkcolor=blue,hyperindex]{hyperref}
\usepackage{braket}
\usepackage{float}
\usepackage{overpic}
\usepackage{bm}
\usepackage{array}
\usepackage{booktabs}
\usepackage{appendix}
\usepackage{tabularx}
\usepackage{cases}
\usepackage{multirow}
\usepackage{mathtools}
\usepackage{soul}
\usepackage{amsmath}
\usepackage{mathrsfs}

\def\bs{\mathbf{S}}

\begin{document}

\title{Spin order and spin excitation spectra of spin-1/2 tetramer chains}
\author{Junli Li}
\affiliation{State Key Laboratory of Optoelectronic Materials and Technologies, Guangdong Provincial Key Laboratory of Magnetoelectric Physics and Devices, Center for Neutron Science and Technology, School of Physics, Sun Yat-Sen University, Guangzhou 510275, China}
\author{Jun-Qing Cheng}
\affiliation{School of Physical Sciences, Great Bay University, Dongguan 523000, China, and Great Bay Institute for Advanced Study, Dongguan 523000, China}
\author{Trinanjan Datta}
\email[Corresponding author:]{tdatta@augusta.edu}
\affiliation{Department of Physics and Biophysics, Augusta University, 1120 15$^{th}$ Street, Augusta, Georgia 30912, USA}
\author{Dao-Xin Yao}
\email[Corresponding author:]{yaodaox@mail.sysu.edu.cn}
\affiliation{State Key Laboratory of Optoelectronic Materials and Technologies, Guangdong Provincial Key Laboratory of Magnetoelectric Physics and Devices, Center for Neutron Science and Technology, School of Physics, Sun Yat-Sen University, Guangzhou 510275, China}
\begin{abstract}
We investigate the spin dynamics of a 1D spin-1/2 Heisenberg tetramer chain. Employing a combination of density matrix renormalization group, quantum renormalization group, and perturbation theory techniques, we compute the energy levels and the quantum phase diagram, analyze the phase transitions, and evaluate the dynamical structure factor (DSF), which is related to the inelastic neutron scattering (INS) spectrum and the resonant inelastic x-ray scattering (RIXS) spectrum of fractionalized and collective excitations. Our calculations suggest that the chain can transition between a hidden $Z_2\times Z_2$ discrete symmetry preserving tetramer phase and a Haldane phase with non-vanishing string order that breaks the hidden symmetry. These two gapped phases are intervened by an intermediate deconfined quantum critical state comprising of free spins and three-site doublets, which is a gapless critical phase with deconfined spinons that can be detected either by INS or $L$-edge RIXS. We find that the tetramer chain can support fractionalized (spinon) and collective (triplon and quinton) excitations. In the ferromagnetic intra-tetramer limit, the chain can support a quinton excitation which has a five-fold degenerate excited state. String order parameter calculations suggest CuInVO$_5$ should be in a Haldane-like phase whose INS or $L$-edge RIXS spectrum can support observable single-spin excitations including triplon and quinton excitations. We also identify possible two-particle excitations (two-singlon, two-triplon, triplon-quinton, and two-quinton excitations) resulting from the double spin-flip effect in the $K$-edge RIXS spectrum.
\end{abstract}


\maketitle

\section{Introduction}\label{sec:Introduction}




The exploration of topological matter has unveiled exotic quantum phases characterized by non-local entanglement and robust boundary phenomena~\cite{Haldane1983464,KatsuraPhysRevLett.104.066403,ChenPhysRevX.8.041028}. In two-dimensional honeycomb lattices~\cite{Owerre_2016,ChenPhysRevX.8.041028} and frustrated magnets~\cite{KatsuraPhysRevLett.104.066403,OwerrePhysRevB.95.014422,LeePhysRevB.97.180401}, topological magnons emerge as collective spin excitations with non-trivial Berry curvature~\cite{LeePhysRevB.97.180401,ZhangPhysRevB.103.174402,SheikhiPhysRevB.104.045139}. Such topological excitations are intrinsically linked to the concept of topological order, a quantum phase beyond Landau symmetry-breaking paradigms, defined by long-range entanglement and ground state degeneracy sensitive to global topology~\cite{HungPhysRevLett.114.076401,LuPhysRevLett.125.116801,WangPRXQuantum.6.010314}.


In one-dimensional (1D) quantum spin chains, the spin order and spin dynamical properties display a rich variety of quantum phenomena~\cite{JafariPhysRevB.76.014412,ChengNPJQM2022,ChengNPJQM2024,PrabhakaPhysRevB.111.205106,LiPhysRevB.111.024404}. These systems can harbor fractionalized excitation such as spinon ~\cite{BalentsPhysRevB.101.020401,WangPhysRevB.105.184429,ChengNPJQM2022,LiPhysRevB.111.024404} or serve as a material platform for symmetry-protected topological (SPT) phases~\cite{WenPhysRevB.80.155131,FidkowskiPhysRevB.83.075103,ShapourianPhysRevLett.118.216402}, which possess string order~\cite{GongPhysRevB.78.104416}. In this context, the subtle distinction between integer versus half-integer spin chains is captured by the Haldane conjecture~\cite{Haldane1983464,AffleckPhysRevLett.59.799}. It states that the ground state of an integer spin chain is characterized by a finite energy (Haldane) gap that appears in the excitation spectrum. Additionally, it has now been realized that the Haldane phase (an example of a SPT phase) can support topologically protected edge states. In contrast, the isotropic half-integer-spin chains tend to have gapless ground states~\cite{ScheiePhysRevB.103.224434}, which is critical, and lack the topological features associated with the Haldane phase of spin-1 chains~\cite{Haldane1983464,WhitePhysRevB.77.134437,SharmaPhysRevB.111.064404}. The spin-$1/2$ chains exhibit gapless excitations and long-range quantum correlations characteristic of a Luttinger liquid~\cite{giamarchi}. Ladders of coupled spin-$1/2$ chains ~\cite{HakobyanPhysRevB.63.144433} and dimerized spin chains~\cite{HidaPhysRevB.45.2207} can support a Haldane-like phase while the isotropic spin-$1/2$ chain does not exhibit a Haldane phase. Theoretically, the presence of a Haldane-like phase~\cite{WenPhysRevB.80.155131} can be detected using topological string order~\cite{HidaPhysRevB.45.2207,SuPhysRevB.78.104416}, which is captured by a string order parameter~\cite{HidaPhysRevB.45.2207}. 

Gapless fractionalized spin excitations and collective gapped high-energy spin excitations have been proposed to be observed in the trimer Heisenberg antiferromagnetic chain~\cite{ChengNPJQM2022,ChengNPJQM2024,LiPhysRevB.111.024404}, while the dimer Heisenberg antiferromagnetic chain only produces collective high-energy spin excitations. Therefore, gapped high-energy excitations are observed when the spin chains become dimerized~\cite{ChitraPhysRevB.52.6581,FurukawaPhysRevB.86.094417,HanPhysRevB.96.125105} and trimerized~\cite{ChengNPJQM2022,ChengNPJQM2024,LiPhysRevB.111.024404,BeraNatComm2022}. Similar to the dimerized and the trimerized cases, a spin-$1/2$ tetramerized spin chain can exhibit a Haldane-like phase under appropriate conditions~\cite{SuPhysRevB.78.104416}. The Haldane-like phase can be investigated experimentally~\cite{LiPhysRevLett.127.263004,Songsciadv.aao4748,Songsciadv.aao4748,Léséleucscience.aav9105} and theoretically~\cite{SuPhysRevB.78.104416,WenPhysRevB.80.155131,PollmannPhysRevB.86.125441,GongPhysRevB.78.104416,HidaPhysRevB.45.2207}. Dimerization can create an effective spin-1 system, thereby allowing for the possibility of a Haldane-like phase. As described later in the manuscript, such gapped phases can generate high-energy collective excitations (triplon, quarton, and quinton excitations), and multi-particle excitations (e.g., two-triplon and triplon-quarton excitations). The nature of these excitations, the associated phases and phase transitions, and the multi-particle fractionalized and collective excitations can be investigated using the string order parameter~\cite{HidaPhysRevB.45.2207,GongPhysRevB.78.104416}. 

In a dimer Heisenberg antiferromagnetic chain, the energy states split into the ground and excited triplet states. The transition from the singlet ground state to the excited singlet, triplet, and quintet states, creating high-energy excitations. For a trimer Heisenberg antiferromagnetic chain, the energy state for a trimer can be excited from the ground state to the excited doublet or quartet state. The corresponding collective spin excitations are doublons and quartons, respectively~\cite{ChengNPJQM2022,ChengNPJQM2024,LiPhysRevB.111.024404}. However, currently there is no study that investigates how these excitations can emerge and, most importantly, what the associated experimental signatures are. Additionally, the energy of these excitations lies above the range of the spinons and the magnons~\cite{SuPhysRevB.78.104416,WenPhysRevB.80.155131,MaPhysRevX.13.031016,LiPhysRevLett.127.263004,Songsciadv.aao4748,PollmannPhysRevB.86.125441}. This is because the energy levels for the high-energy excitations are dependent on the intrinsic Hilbert space of a single dimer, trimer, or tetramer unit, while the energy levels for the spinon and the magnon are a consequence of the average exchange interactions of the spin chain.

Excitations in a correlated electronic system span from low to high energy, including multi-spin excitations. Thus, one may need to utilize multiple spectroscopic tools to investigate the full range of the excitation spectrum. Typically, inelastic neutron scattering (INS) is utilized to investigate magnetic properties~\cite{GaoPhysRevB.109.L020402}. Although INS is effective at detecting single-spin excitations quite accurately, accessing the higher-order spin excitation modes that arise in a spin chain can be difficult. Thus, we analyze the origins of the 1D tetramer spin chain phases and compute the resonant inelastic x-ray scattering (RIXS) spectrum, an experimental technique that can adequately probe high-energy excitations of a 1D quantum spin chain~\cite{NoceraPhysRevE.94.053308,SchlappaNatComm2018,NoceraSciRep2018,KumarPhysRevB.102.075134,NagNatComm2022,Schmiedinghoff2022}. Note, our DMRG simulations~\cite{FortePhysRevB.83.245133,NagNatComm2022,GaoPhysRevB.109.L020402,LiPhysRevB.111.024404} can be applied to analyze both INS and RIXS data. In neutron spectroscopy, one detects a single-spin excitation process, which is similar to $L$-edge RIXS~\cite{FortePhysRevB.83.245133,NagNatComm2022,LiPhysRevB.111.024404}. Thus, 
the effects of the dynamical structure factor (DSF) can be detected by both INS~\cite{GaoPhysRevB.109.L020402} and RIXS~\cite{NagNatComm2022}. The two-particle excitation process, which can be detected in $K$-edge RIXS~\cite{FortePhysRevB.77.134428,IshiiPhysRevB.112.195132}, is described by a four-spin correlation function. 

In this article, we investigate the excitation spectrum of the 1D spin-$1/2$ Heisenberg tetramer chain. Utilizing a combination of quantum renormalization group~\cite{KargarianPhysRevA.77.032346} and perturbation theory~\cite{ChengNPJQM2022}, we unravel the fractionalized and collective excitation spectra of the tetramer chain. We utilize density matrix renormalization group (DMRG) to compute the phase diagram, single-spin excitation spectra, and the $L$-edge and $K$-edge resonant inelastic x-ray scattering (RIXS) spectra using the Krylov-space correction vector (CV) method in the DMRG framework~\cite{NoceraPhysRevE.94.053308,NoceraSciRep2018}. Note, by assuming that the tetramer chain has a relatively small $J/\Gamma$ ($J$ is the exchange interaction in the system and $\Gamma$ is the core-hole energy broadening), which results in a short core-hole lifetime, we compute both the $L$-edge and $K$-edge RIXS intensity response functions within the ultrashort core-hole lifetime (UCL) expansion framework~\cite{Brink_2007,FortePhysRevB.83.245133,NagNatComm2022}. Crucially, our work establishes INS and RIXS as pioneering tools for probing topological order in quantum magnets, which is the first application of this technique to directly resolve symmetry-protected topological (SPT) phases in a quasi-1D material. Our analysis of the quantum phase transition, based on the string order parameter, suggests the possibility of a phase transition between a trivial tetramer phase (with zero string order parameter) and a SPT Haldane phase (with non-vanishing string order). The boundary between these two phases is a gapless quantum critical state with deconfined spinons~\cite{Senthilscience.1091806,JiangPhysRevB.99.075103,RobertsPhysRevB.99.165143,HuangPhysRevB.100.125137,Pronks53y-qmr4,YangPhysRevE.104.064121,LiPhysRevB.107.085130}. The spinon in the spin-1/2 tetramer chain appears when the system is in the deconfined quantum critical point (DQCP), which is called the deconfined spinon state in our work. The tetramer chain is able to host a collection of single-particle excitations (spinon, triplon, and quinton excitations), which can be detected both by the INS and $L$-edge RIXS spectra. Numerous two-particle excitations (two-triplon, triplon-quinton, and two-singlon excitations) can be realized in $K$-edge RIXS. Based on string order parameter calculations, we find that CuInVO$_5$ is a candidate material that can be in the Haldane phase~\cite{RejaPhysRevB.99.134420}. Remarkably, our DMRG simulations of CuInVO$_5$ not only confirm the existence of the Haldane phase but also demonstrate the unique capability of RIXS to detect its collective excitations, which is a breakthrough in linking spectroscopic signatures to SPT order. We find that the RIXS spectra of CuInVO$_5$ can support the triplon and quinton excitations. This constitutes the first unambiguous identification of collective excitations in a candidate SPT material via x-ray scattering techniques. Note, both INS and RIXS can only detect high-energy gapped excitations in a spin-1/2 tetramer chain when the system is in the trivial tetramer phase and the Haldane phase. However, additional low-energy gapless modes appear in the excitation spectra when the system is in the deconfined spinon state, which is the boundary of the tetramer phase and the Haldane phase. In the rest of the article, we state our results and discuss our findings. This is followed by an explanation of our numerical and analytical methods.

\begin{figure*}[t]
\centering
\centerline{\includegraphics[width=17.2cm]{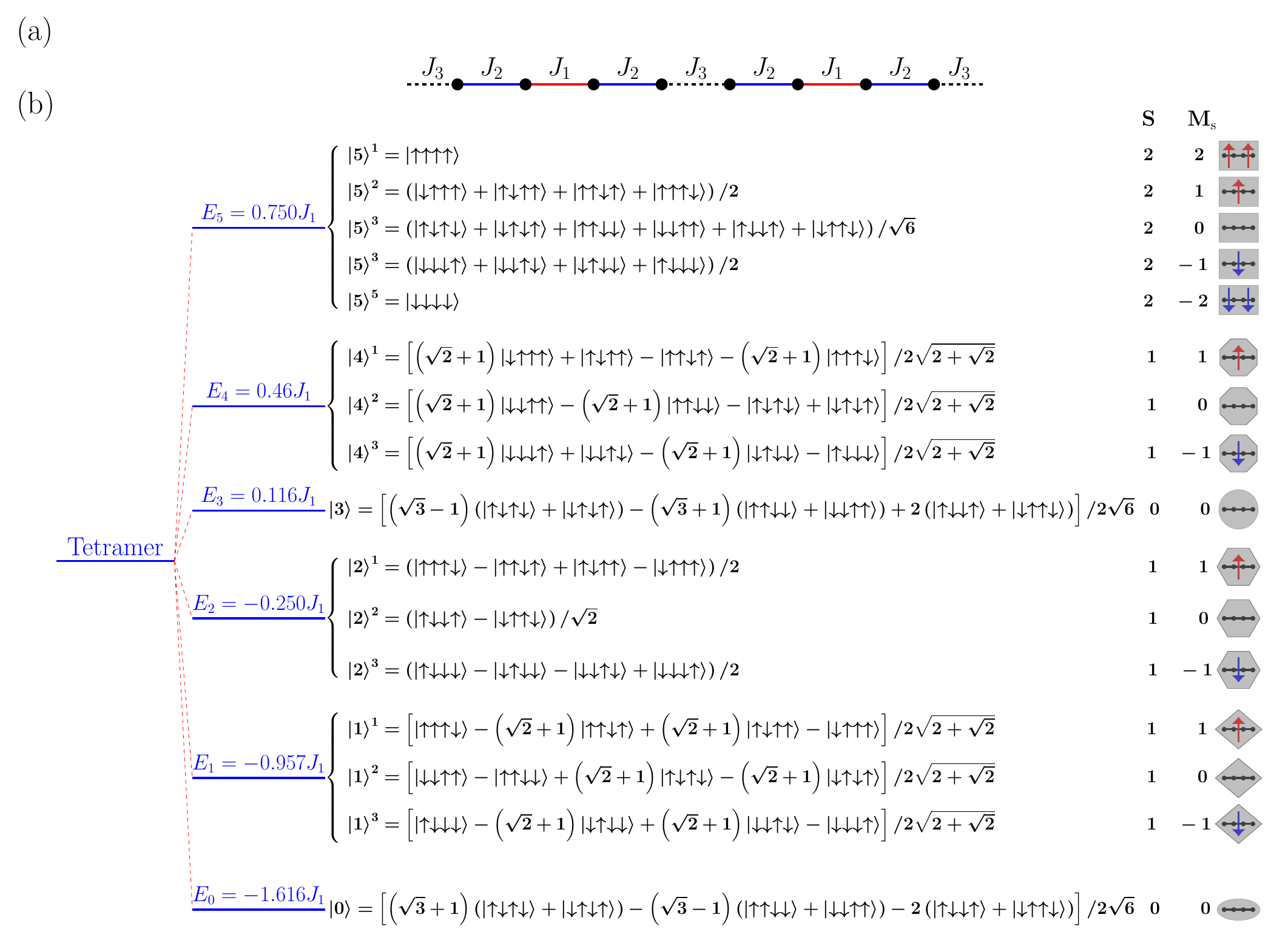}}
\caption{Tetramer spin chain with its interaction definitions, the energy states of a single tetramer unit, and the symbols used to represent the various tetramer states. (a) A tetramer spin chain includes intra-tetramer exchange interactions $J_1$ and $J_2$. The corresponding relative coupling strength is defined by $\alpha=J_2/J_1$. The inter-tetramer exchange interaction is given by $J_3$. The relative inter-tetramer coupling strength is defined as $\beta=J_3/J_1$. (b) The energy levels and the corresponding wave functions for a single tetramer unit are computed using exact diagonalization. The symbol $E_\epsilon~(\epsilon=0\dots 5$) denotes the lowest to the highest energy levels, respectively. The ground state $\ket{0}$ is represented using an ellipse. The excited states $\ket{\epsilon},~\epsilon=0\dots 5$, are denoted using a diamond, a hexagon, a circle, an octagon, and a rectangle, respectively.}
\label{fig:fig1}
\end{figure*}

\begin{figure}[h]
\centering
\centerline{\includegraphics[width=8.7cm]{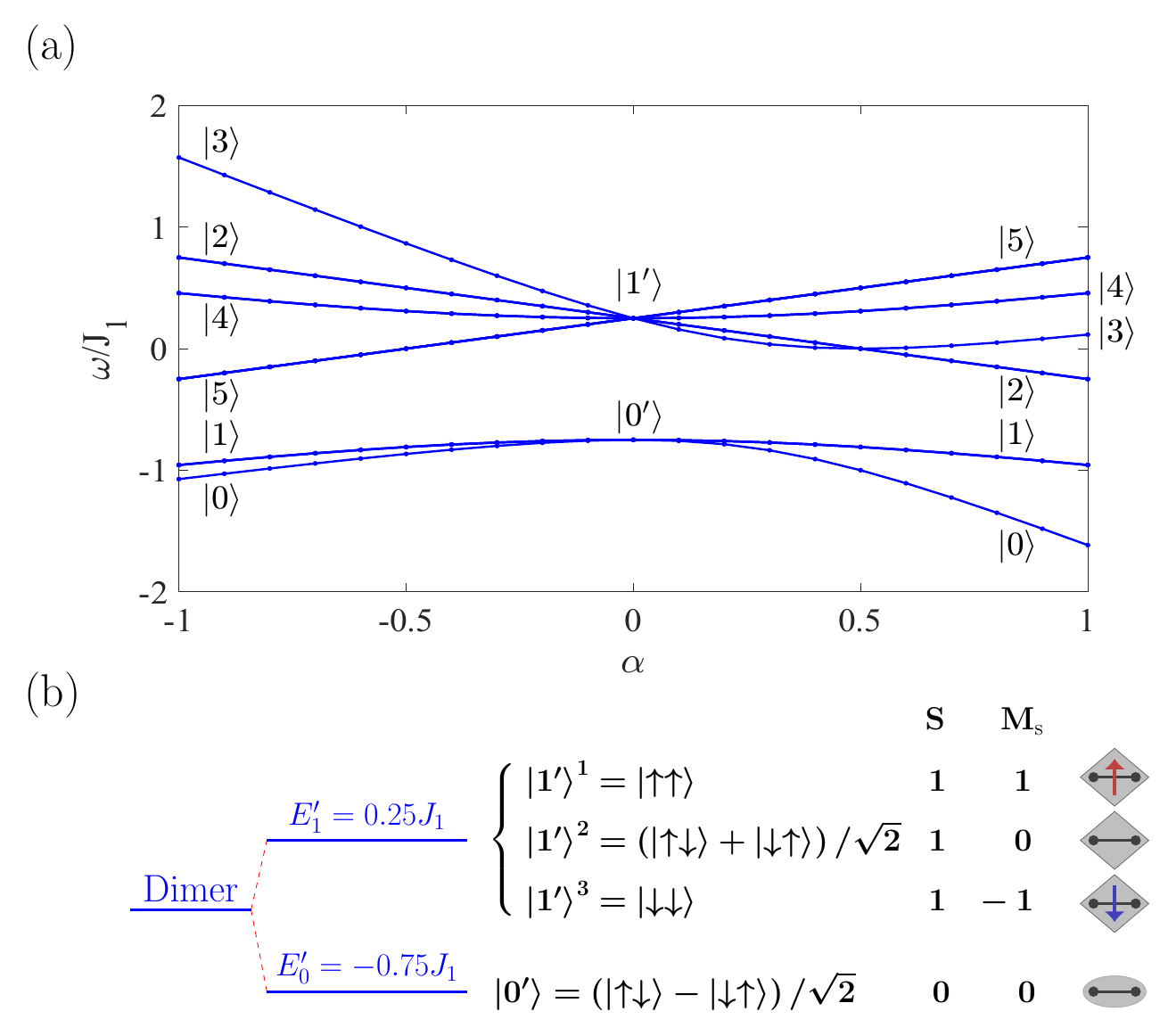}}
\caption{Energy level diagrams of a tetramer unit for various intra-tetramer coupling strengths $\alpha$ and the energy states of a dimer unit. (a) The energy levels of a single tetramer unit with $\alpha$ ranging from $-1$ to $1$. The energy states $\ket{\epsilon}, \epsilon=0\dots 5$ belong to the tetramer states. The energy states $\ket{0^{\prime}}$ and $\ket{1^{\prime}}$ belong to the dimer. (b) The ground state and the excited state energy of a dimer unit, computed using exact diagonalization, are denoted by $E^{\prime}_0$ and $E^{\prime}_1$, respectively. The ellipse and the diamond symbols refer to the ground and the excited states, respectively. Note that in Fig.~\ref{fig:fig1}(b), four black dots denote a tetramer state. While the dimer state is denoted using two black dots.}
\label{fig:fig2}
\end{figure}

\section{Results}\label{sec:Results}

\subsection{Model}\label{sec:Model}
The spin-1/2 Heisenberg tetramer chain Hamiltonian of length L with $N = \text{L}/4$ tetramers is given by \begin{equation}
\begin{split}
\label{eq:ham}
\hat{H}=&\sum^N_{n=1}\left[J_2\left(\hat{\bs}_{4n-3}\cdot\hat{\bs}_{4n-2}+\hat{\bs}_{4n-1}\cdot\hat{\bs}_{4n}\right)\right. \\
&\left.+J_1\hat{\bs}_{4n-2}\cdot\hat{\bs}_{4n-1}+J_3\hat{\bs}_{4n}\cdot\hat{\bs}_{4n+1}\right].
\end{split}
\end{equation}
The spin operator $\hat{\bs}_j$ at site $j$ spans over the tetramer site indices as shown in Eq.~\ref{eq:ham}. The exchange couplings $J_1$ and $J_2$ denote intra-tetramer exchange interactions, while $J_3$ represents the inter-tetramer exchange interaction. A schematic representation of this model is shown in Fig.~\ref{fig:fig1}(a). We define the relative intra-tetramer and inter-tetramer coupling strengths as $\alpha=J_2/J_1$ (with $\alpha\in [-1,1]$)  and $\beta=J_3/J_1$ (with $\beta\in [0,1]$), respectively. In Fig.~\ref{fig:fig1}(b), we display the energy levels of a single tetramer unit ($\alpha$$\neq$0, $\beta$=0). The energy scale of $J_1$ is 20.7 meV for the tetramer compound CuInVO$_5$ in the \emph{Ground state and excited states for CuInVO$_5$} section~\cite{RejaPhysRevB.99.134420}.

The energy levels of the tetramer, computed using exact diagonalization, are shown in the $\omega/J_1$ (energy) vs $\alpha$ plot (see Fig.~\ref{fig:fig2}(a)). The spin-1/2 tetramer Heisenberg chain has six energy levels when the intra-tetramer coupling $\alpha$ is non-zero. The corresponding energy states include two singlet states, three triplet states, and one quintet state. These energy states rearrange based on the value of $\alpha$. A single spin-flip causes the ground state to transition to a triplet or a quintet state, which supports the triplon and the quinton excitation, respectively. Six different energy levels $|\epsilon\rangle$, where $\epsilon=0\dots 5$, are presented in a spin-1/2 tetramer unit. When $\alpha=1$, the energy states are arranged in an ascending order beginning with $|0\rangle$. The energy levels and the corresponding wave functions for $\alpha=1.0$ are presented in Fig.~\ref{fig:fig1}(b). The ground state $\ket{0}$ and the third excited state $\ket{3}$ are singlets. The excited states $\ket{1}$, $\ket{2}$, and $\ket{4}$ are triplets. The highest excited state $\ket{5}$ is a quintet. The energy expressions $E_0$ to $E_5$ can be written as \begin{equation}
\label{eq:E0E5}
\begin{split}
E_0\left(J_1,J_2\right)&=\frac{1}{8}\left(-\tilde{J_1}-2\tilde{J_2}-2\sqrt{\tilde{J_1}^2+4\tilde{J_2}^2-2\tilde{J_1}\tilde{J_2}}\right), \\
E_1\left(J_1,J_2\right)&=\frac{1}{8}\left(-\tilde{J_1}-2\sqrt{\tilde{J_1}^2+\tilde{J_2}^2}\right), \\
E_2\left(J_1,J_2\right)&=\frac{1}{8}\left(\tilde{J_1}-2\tilde{J_2}\right), \\
E_3\left(J_1,J_2\right)&=\frac{1}{8}\left(-\tilde{J_1}-2\tilde{J_2}+2\sqrt{\tilde{J_1}^2+4\tilde{J_2}^2-2\tilde{J_1}\tilde{J_2}}\right), \\
E_4\left(J_1,J_2\right)&=\frac{1}{8}\left(-\tilde{J_1}+2\sqrt{\tilde{J_1}^2+\tilde{J_2}^2}\right), \\
E_5\left(J_1,J_2\right)&=\frac{1}{8}\left(\tilde{J_1}+2\tilde{J_2}\right),
\end{split}
\end{equation}
where $\tilde{J_1}=J_1+\left|J_1\right|$ and $\tilde{J_2}=J_2+\left|J_2\right|$. 

When $\alpha=0$, the tetramer unit transforms to a dimer system. These energy levels are shown in the upper half of Fig.~\ref{fig:fig2}(a). The ground state $\ket{0}$ and the first excited state $\ket{1}$ of the tetramer unit combine to form the dimer ground state singlet $\ket{0^{\prime}}$. The other tetramer excited states $\ket{2}$, $\ket{3}$, $\ket{4}$, and $\ket{5}$ coalesce together to form the dimer excited triplet state $\ket{1^{\prime}}$. The energy levels are rearranged when $\alpha$ is negative (see Fig.~\ref{fig:fig2}(a)). We note that the energy level for $\ket{3}$ becomes lower than the energy level of $\ket{2}$ when $\alpha$ is in the range $\alpha\in [0,0.5]$. In Fig.~\ref{fig:fig2}(b), we show the energy and the corresponding energy states when $\alpha=0$. It is indicated that the tetramer unit establishes a dimer state with only two energy states when intra-tetramer interaction $J_2$ is close to zero. The ground state $\ket{0^{\prime}}$ is a singlet, and the first excited state $\ket{1^{\prime}}$ is a triplet. Here, we show the energy states of the spin-1/2 tetramer chain when $\alpha=1$ to motivate the discussion for the rest of the article. The tetramer states of the compound CuInVO$_5$, which are reordered when $\alpha < 1$, are discussed in the \emph{Ground state and excited states for CuInVO$_5$} section. The above discussion concludes our analysis of the energy and the corresponding energy states of a single tetramer and dimer unit. In the next section, we investigate the behavior of the spin-1/2 chain formed by tetramer units.

\subsection{Quantum phase analysis}\label{sec:Quantum phase analysis}

\begin{figure*}[t]
\centering
\centerline{\includegraphics[width=18.5cm]{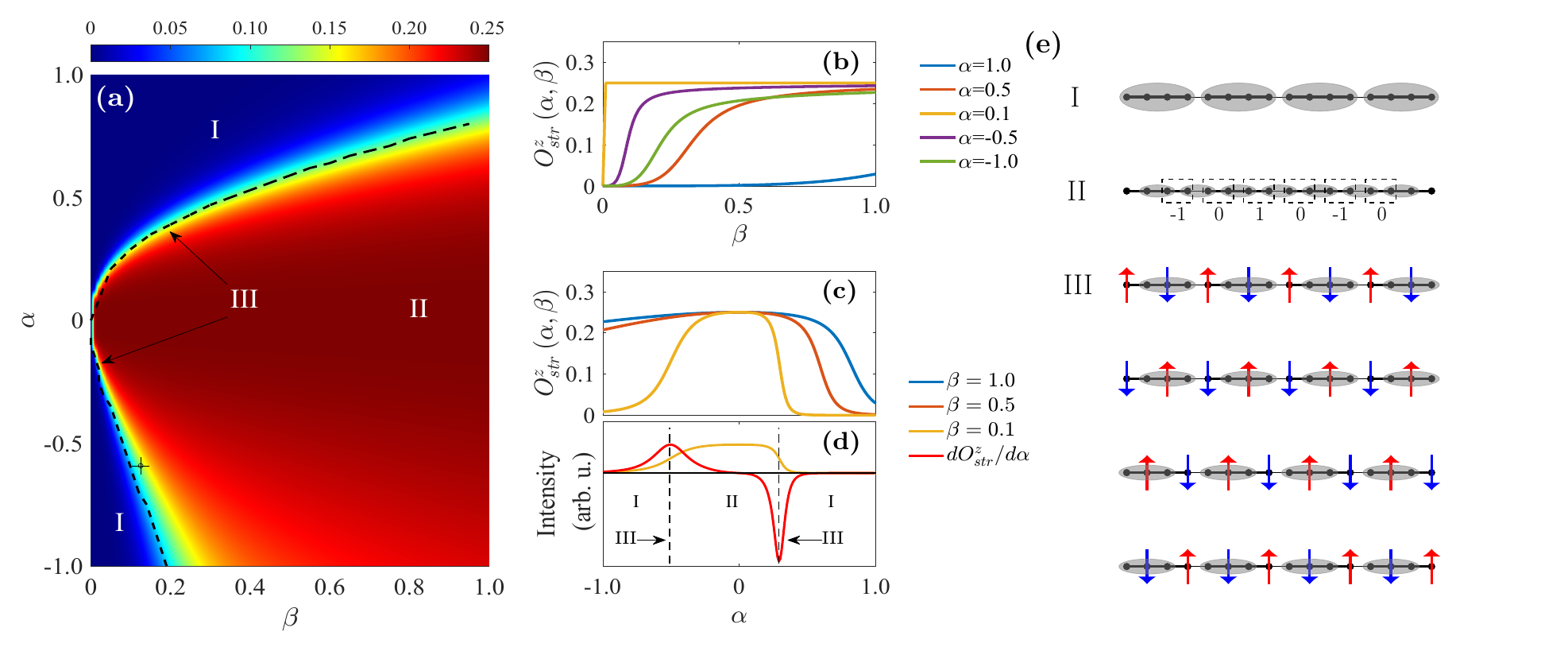}}
\caption{String order parameter and three different kinds of ground states for tetramer spin chains computed using DMRG. (a) String order parameter $O^z_{str}(\alpha,\beta)$ as a function of $\alpha$ and $\beta$. The approximate locations of two different phases are marked by I and II. Zone III indicates the phase transition boundary between the tetramer phase and the Haldane phase. The string order parameter $O^z_{str}(\alpha,\beta)$ with $\alpha\in[-1,1]$ and $\beta\in[0,1]$ captures most of the features of the tetramer system. In Supplementary Note Figure 4, we show the results of the string order parameter for extended ranges of $\alpha$ and $\beta$ values covering $\alpha\in[-55,5]$ and $\beta\in[0,20]$. We have included a discussion on this extended zone in Sec. F of the Supplementary Note. (b) String order parameter for $\beta\in\left[0,1\right]$ and $\alpha=-1.0,-0.5,0.1,0.5$, and $1.0$. (c) String order parameter for $\alpha\in\left[-1,1\right]$ and $\beta=0.1,0.5$, and $1.0$. (d) The first-order derivative $dO^z_{str}/d\alpha$ with $\alpha\in\left[-1,1\right]$ and $\beta=0.1$. (e) The schematic pictures for three different kinds of ground states. Label I represents a state that forms tetramer singlets and is in a tetramer phase. Label II indicates that a string order has been established in the Haldane phase. The dashed boundary between the tetramer phase and the Haldane phase has a ground state that is proliferated by deconfined spinons in a gapless quantum critical state. We identify this boundary with the label III. The transition at this boundary exhibits deconfined quantum criticality~\cite{JiangPhysRevB.99.075103,RobertsPhysRevB.99.165143,HuangPhysRevB.100.125137,Pronks53y-qmr4,YangPhysRevE.104.064121,LiPhysRevB.107.085130}.  We also indicate the location of the compound CuInVO$_5$ (a candidate Haldane-like material) by a cross-hair symbol on the string order parameter phase diagram~\cite{HasePhysRevB.94.174421,RejaPhysRevB.99.134420}.}
\label{fig:fig3}
\end{figure*}

The string order is a measure of the formation and the disappearance of hidden symmetry in systems that do not exhibit conventional magnetic order (such as ferromagnetic or antiferromagnetic). In a dimerized spin chain, the string order reveals whether the underlying phase has a hidden $Z_2\times Z_2$ discrete symmetry or not. When dimerization is introduced, the system can transition to the non-trivial Haldane phase. The string order parameter is enhanced if the dimerization is strong. However, for weak dimerization, the value of the string order parameter may remain small and close to zero, signaling a phase that has trivial features. The non-locality of the string order parameter, which captures the underlying hidden order, is in stark contrast to the local nature of dimerization that is realized in the spin chain. The quantum phases of the spin-1/2 dimer ~\cite{HidaPhysRevB.45.2207} have been analyzed using the string order parameter. The tetramer chains have been analyzed using the string order parameter~\cite{GongPhysRevB.78.104416} and spin-spin correlations~\cite{SinghaniaPhysRevB.98.104429}.


Considering the SU(2) symmetry in a spin-1/2 tetramer chain, we adopt the following definition of the string order parameter \begin{equation}
\label{eq:string}
O^z_{str}\left(\alpha,\beta\right)=\lim\limits_{|n-n^{\prime}|\to\infty}\Theta^z_{4n,4n^{\prime}+1},
\end{equation} and compute it using DRMG (see the \emph{Method} section). In the above, the string operator is given by\begin{equation}
\label{eq:Oij}
\Theta^z_{4n,4n^{\prime}+1}=-\left\langle S^z_{4n}e^{i\pi\left(S^z_{4n+1}+S^z_{4n+2}+~\cdots~+S^z_{4n^{\prime}-1}+S^z_{4n^{\prime}}\right)}S^z_{4n^{\prime}+1}\right\rangle.
\end{equation} In the above equation the spin operator is $S^z_j=e^{i\pi S^z_j}/2i$. The string order parameter behavior can be predicted by tracking the hidden symmetry, which is found by a nonlocal unitary dual transformation~\cite{HidaPhysRevB.45.2207,SuPhysRevB.78.104416}. This operator has been defined to ensure that all the exchange interactions $J_1$, $J_2$, and $J_3$ are considered in the tetramer chain. Next, according to the standard Kramers-Wannier dual transformation formalism~\cite{KramersPhysRev.60.252}, the spin-1/2 tetramer chain Hamiltonian in Eq.~\ref{eq:ham} is transformed into the following expression\begin{equation}
\label{eq:hamdual}
\begin{split}
\tilde{H}/J_1&=\beta\sum_n\left(\sigma^z_{2n}\sigma^z_{2n+1}+\tau^z_{2n}\tau^z_{2n+1}-\sigma^z_{2n}\sigma^z_{2n+1}\tau^z_{2n}\tau^z_{2n+1}\right) \\
             &+\sum_n\left(\sigma^x_{2n}-\beta\sigma^x_{2n+1}\right)+\sum_n\left(\tau^x_{2n}-\beta\tau^x_{2n+1}\right) \\
             &+\sum_n\left(\beta\sigma^x_{2n+1}\tau^x_{2n+1}-\sigma^x_{2n}\tau^x_{2n}\right),
\end{split}
\end{equation} where the Pauli matrices $\sigma$ and $\tau$ represent the spins on integer and half-integer site positions, respectively. The hidden $Z_2\times Z_2$ discrete symmetry is found due to the rotational invariance of $\pi$ on the $x$ axis of the $\sigma$ spins and $\tau$ spins, where the $\sigma$ and $\tau$ should be Pauli matrices that include $\sigma^{\mu}$ and $\tau^{\mu}, \mu = x,y,z$. According to the dual transformation, the string operator $\Theta^z_{4n,4n^{\prime}+1}$ is the matrix product of the $x$ component of $\sigma$ spins $U\Theta^z_{4n,4n^{\prime}+1}U^{-1}=-\otimes^{2n^{\prime}+1}_{i=2n+1}\sigma^x_i$, where $U$ is the dual transformation operator for the entire spin-1/2 tetramer chain. This means that the string order parameter $O^z_{str}\left(\alpha,\beta\right)$ should vanish in a trivial phase but remain non-zero in a Haldane phase when the $Z_2\times Z_2$ symmetry is broken.

We compute the string order parameter $O^z_{str}\left(\alpha,\beta\right)$ over the parameter space $\alpha\in[-1,1]$ and $\beta\in[0,1]$ using DMRG (see the \emph{Method} section). The results are displayed in Fig.~\ref{fig:fig3}, where the blue (red) color in Fig.~\ref{fig:fig3}(a) represents the opposing limits of a trivial (Haldane) phase. In the following discussion, we will first identify the different phases arising in the tetramer chain. Next, we will present a conceptual picture of each phase, followed by the various excitations that are supported in that phase. Upon inspecting Fig.~\ref{fig:fig3}(a), we notice that the minimum value of $O^z_{str}\left(\alpha,\beta\right)$ is located at the top and bottom left corners of the plot, which are the regions where $\left|\alpha\right|\approx 1$ and $\beta \rightarrow 0$ (limit of isolated tetramer units). For $\alpha>0$, the trivial phase spans a larger parameter space compared to $\alpha<0$. These regions are classified as the tetramer phase (trivial phase). The maximum value of the string order is located at the right edge of the diagram where $\left|\alpha\right|\approx 0$ and $\beta=1.0$. For this choice, the chain is in the Haldane phase~\cite{HidaPhysRevB.45.2207}. We have confirmed that this is a second-order quantum phase transition (see Sec.~A and Supplementary Figure~1 in the Supplementary Note). The spin excitation spectra indicate that the spin-1/2 tetramer chain generates gapless excitations with the parameter sets between the tetramer phase and the Haldane phase. This intermediate zone is labeled as III, which is described as a gapless critical phase of the system where the spinons are deconfined.

In Fig.~\ref{fig:fig3}(b), the blue curve shows that the string order parameter reaches its minimum when $\alpha=1.0$ for all values of $\beta$. The curves for $\alpha=-1.0,-0.5,0.1$, and $0.5$ all converge to a similar value when $0.5<\beta<1$. In the region between $0<\beta<0.5$, the curve for $\alpha=0.5$ decreases the most significantly compared to the curves for $\alpha=0.1,-0.5,-1.0$. While the curve for $\alpha=-1.0$ decays faster than the curve for $\alpha=-0.5$ as $\beta$ decreases to zero. The curve for $\alpha=0.1$ remaining at the same value as $\beta>0$ shows a sudden drop at $\beta=0$ and vanishes. In Fig.~\ref{fig:fig3}(c), the curve for $\beta=1.0$ has the highest value, while the curve for $\beta=0.1$ has the lowest value. As the $\alpha$ increases to zero, all the curves reach the same values at 0.25. And the curves for $\beta=0.1$ and $\beta=0.5$ shrink rapidly at around $\alpha>0.2$ and $\alpha>0.4$, respectively. They become zero when $\alpha=1.0$. While the curve for $\beta=1.0$ drops at about $\alpha>0.5$ and still stays a finite value when $\alpha=1.0$. To locate the critical point of phase transition between the tetramer phase and the Haldane phase, we calculate the string order parameter $O^z_{str}$. As $\beta$ increases from zero, see Fig.~\ref{fig:fig3}(b), both for $\alpha >0$ and $\alpha <0$,  a larger absolute value of $\alpha$ always results in a slower increase of string order, indicating that the tetramer singlet phase is expanded. In Fig.~\ref{fig:fig3}(c), as $\alpha$ approaches zero, all string order curves reach the same values at $0.25$. A smaller value of $\beta$ leads to a narrower Haldane phase.

In Fig.~\ref{fig:fig3}(e), a schematic picture is drawn to indicate the ground state of the tetramer phase, the Haldane phase, and the intermediate gapless critical deconfined spinon state. The tetramer phase represents a ground state where all the tetramer units form singlets along the chain. The tetramer phase, which is a trivial phase where all the sites are included in the tetramer singlets, has the lowest string order parameter value. In the Haldane phase, a string order is formed in the tetramer chain and gives a nonzero value for the string order parameter $O^z_{str}(\alpha,\beta)$. The gapless critical deconfined spinon state III is an intermediate state, where three-site doublets with $S=1/2$ effective spins form in the tetramer system. While one spin in a tetramer unit is excluded from the three-site doublet. The string order parameter in the the gapless critical deconfined state III has intermediate values between the values of the tetramer phase and the Haldane phase. It can be deduced that the intrinsic high-energy excitations for the tetramer system exist in the tetramer phase. The tetramer chain generates gapped excitations from the Haldane phase, which is the typical Haldane phase. While the gapless excitations exist in the deconfined spinon state III.

\begin{figure}[t]
\centering
\centerline{\includegraphics[width=8.5cm]{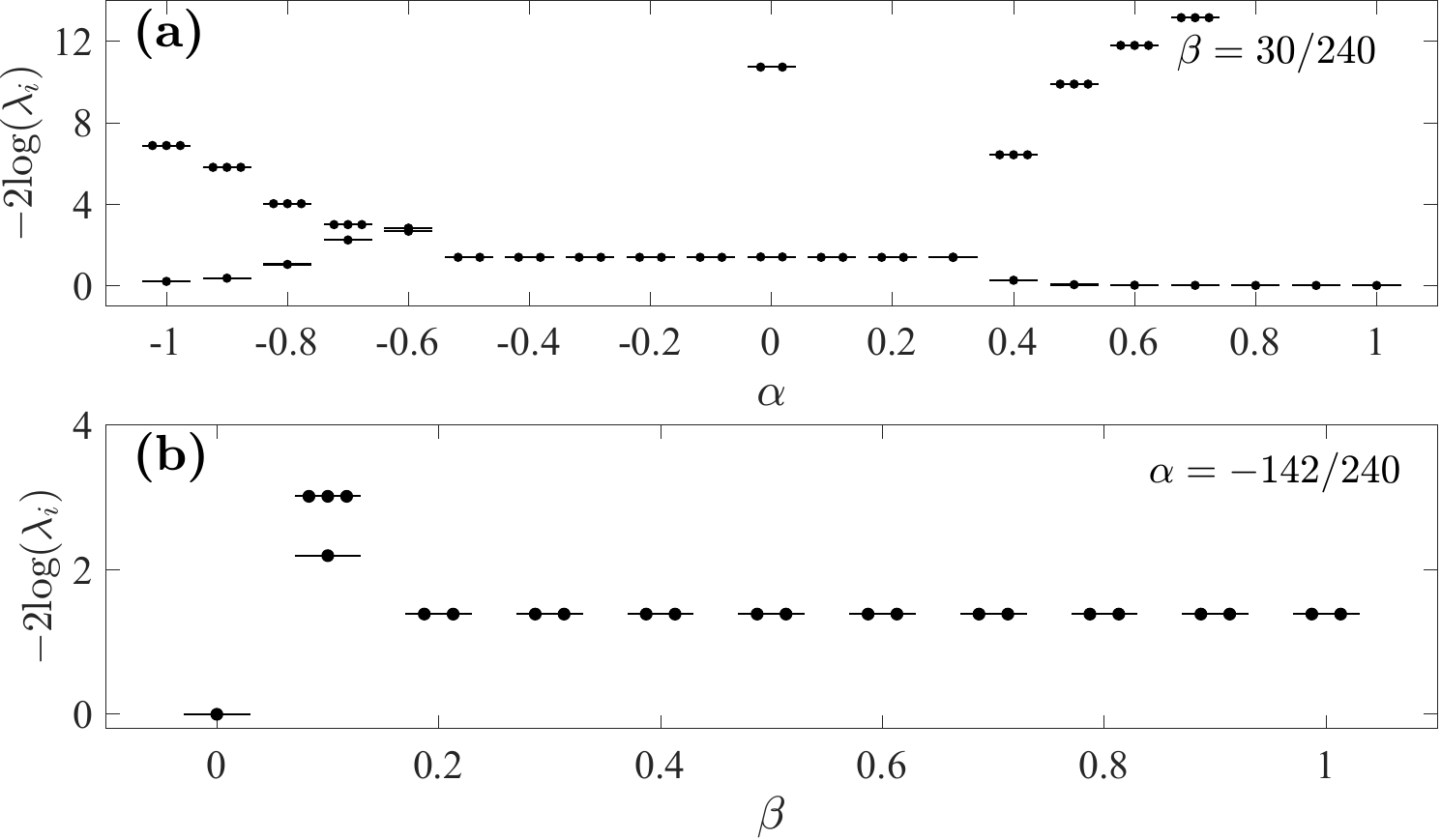}}
\caption{Variation in the entanglement spectrum of the spin-1/2 tetramer chain as the intra- and inter-tetramer couplings are varied. (a) Schmidt weights under different intra-tetramer interaction $\alpha$ with $\beta=30/240$. (b) Schmidt weights under different inter-tetramer interaction $\beta$ with $\alpha=-142/240$.}
\label{fig:fig4}
\end{figure}

To analyze the presence of the Haldane phase in the tetramer chain, we present the entanglement spectra calculation for various $\alpha$ and $\beta$ values. In Fig.~\ref{fig:fig4}, we plot the entanglement spectrum to investigate the variation of the $\alpha$ (intra-tetramer) and the $\beta$ (inter-tetramer) coupling, while either one of them is kept constant. The entanglement spectrum is evaluated from the Schmidt decomposition of the ground state of the tetramer chain. The tetramer chain is treated as a bipartite system from the perspective of DMRG calculation. It is composed of two subsystems, which we label as A and B. The density matrix of subsystem A is given by $\hat{\rho}_A=\text{Tr}_{\text{B}}\ket{\varphi}\bra{\varphi}$, where the eigenvalues of $\rho_A$ are $\lambda^2_i$ (the Schmidt weight). These are equivalent to the eigenvalues of the density matrix of subsystem B. Fig.~\ref{fig:fig4}(a) displays the entanglement spectrum of the spin-1/2 tetramer chain with $\alpha$ ranging from -1 to 1. According to the string order parameter diagram, see Fig.~\ref{fig:fig3}, the strength of $O^z_{str}(\alpha,\beta)$ becomes large in the range $\alpha\in(-0.59, 0.32)$. Comparing this to the $\alpha$ variation in Fig.~\ref{fig:fig4}(a), we can conclude that the Schmidt weights are in even degeneracy in this regime. This suggests the existence of the Haldane phase~\cite{PollmannPhysRevB.81.064439}. In the range $\alpha\in[-1, -0.59]$ and $\alpha\in[0.32, 1]$, the Schmidt weights are in odd degeneracy. Comparing this data to Fig.~\ref{fig:fig3}, we can infer that the chain is in the tetramer phase. Next, we investigate the variation with $\beta$ when $\alpha$ is held fixed. In Fig.~\ref{fig:fig4}(b), the Schmidt weights are in even degeneracy in the range $\alpha\in(0.1, 1]$, thereby indicating the presence of the Haldane phase. In the Haldane phase, string order represents strong entanglement and creates 2-spin singlets along the bulk of the chain. However, at both ends of the chain, the spins become unpaired due to the absence of a nearest-neighbor spin on either side, which results in the formation of the edge states. The presence of the edge states is visible in $\langle S^z_j\rangle$ (see Supplementary Figure 2 and Sec.~E in the Supplementary Note). Both the even-degeneracy and the visibility of edge states indicate the existence of the Haldane phase.

\begin{figure*}[hbtp]
\centering
\centerline{\includegraphics[width=17.8cm]{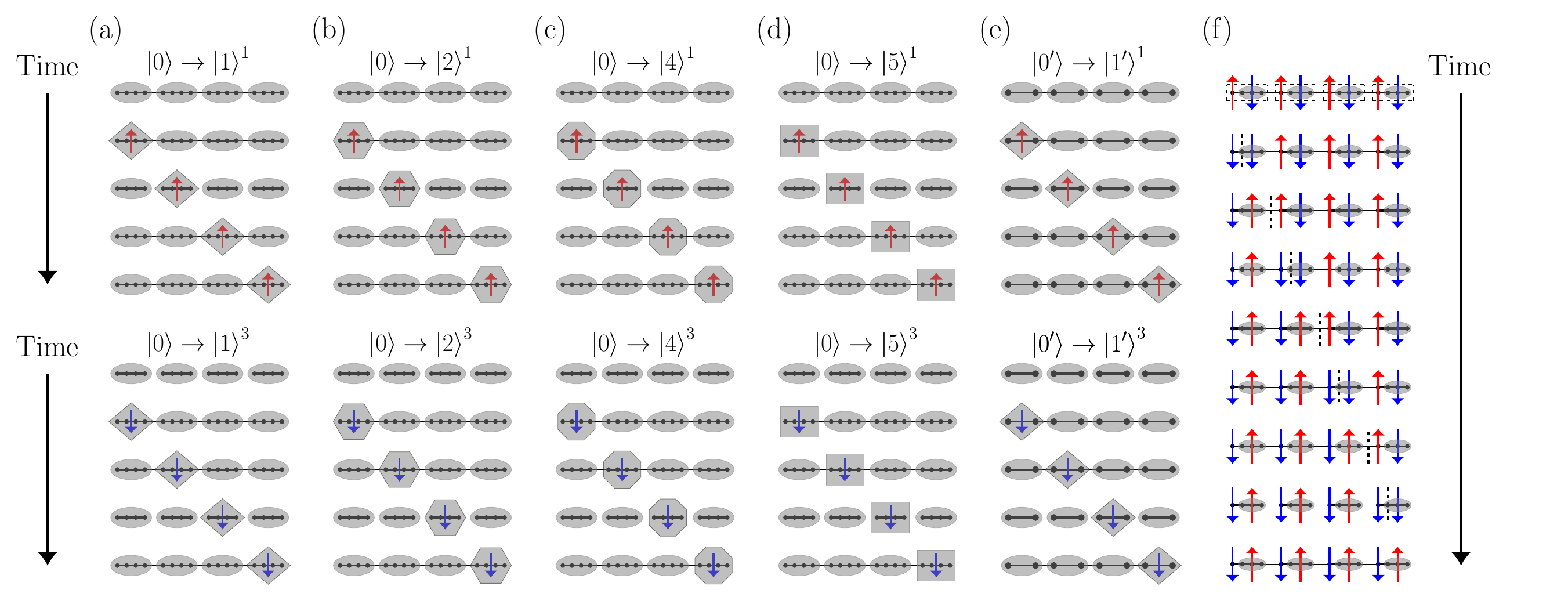}}
\caption{Possible spin excitations of a tetramer spin chain. The red up spins are free spins, while the shaded ellipses with blue down spins are doublets, including three sites. The diamond, hexagon, octagon, and rectangle represent the excited states $\ket{1}$, $\ket{2}$, $\ket{4}$, and $\ket{5}$, respectively, as illustrated in Fig.\ref{fig:fig1} and Fig.\ref{fig:fig2}. The tetramers are surrounded by dashed rectangles. The quantum states $\ket{1}^1$, $\ket{2}^1$, $\ket{4}^1$, $\ket{5}^1$, and $\ket{1^{\prime}}^1$ represent up spins. The down spin states are given by $\ket{1}^3$, $\ket{2}^3$, $\ket{4}^3$, $\ket{5}^3$, and $\ket{1^{\prime}}^3$. (a)~-~(c) Triplon excitations in the tetramer ground state. (d) Quinton excitation in the tetramer ground state. (e) Triplon excitations in the dimer ground state. (f) Spinon excitation in deconfined spinon state III.}
\label{fig:fig5}
\end{figure*}

A visual representation of the possible excitations of the spin-1/2 tetramer chain is shown in Fig.~\ref{fig:fig5}. The high-energy excitations for the tetramer phase are shown in Figs.~\ref{fig:fig5}(a)-(d). In Figs.~\ref{fig:fig5}(a)-(c), the high-energy excitations are triplon excitations excited from tetramer singlets. The quinton excitation, which is only observed when the intra-tetramer coupling $\alpha<0$, is shown in Fig.~\ref{fig:fig5}(d). In Fig.~\ref{fig:fig5}(e), we show the triplon excitation excited from dimer singlets in the Haldane phase. In Figs.~\ref{fig:fig5}(a)-(e), the upper panel is for $\left|\Delta\mathrm{M}_\mathrm{s}\right|=1$ and the lower panel is for $\left|\Delta\mathrm{M}_\mathrm{s}\right|=-1$. Spinon excitations in the deconfined spinon state III are shown in Fig.~\ref{fig:fig5}(f), indicating that the ground state of the system is at the DQCP~\cite{JiangPhysRevB.99.075103,RobertsPhysRevB.99.165143,HuangPhysRevB.100.125137,Pronks53y-qmr4,YangPhysRevE.104.064121,LiPhysRevB.107.085130}. Note, the quinton excitation occurs in the $\left|\Delta\mathrm{M}_\mathrm{s}\right|=1$ mode, which should be detected in INS and $L$-edge RIXS (experimental resolution permitting). We sketch the process by which gapless excitations are created by domain wall propagation, in a potential ground state, near the critical point. In the first row, all the free spins and the effective spins are anti-parallel to each other. A domain wall is created between the spin at the left end of the chain when the first spin flips. Then, the effective spin of the three-site doublet in the tetramer unit at the left end of the chain flips. The domain wall therefore propagates along the tetramer chain. The spin-1/2 tetramer chain supports a gapless excitation because the propagating domain wall flips all the spins and the effective spins when transporting along the chain.  The ground state of the tetramer phase, the Haldane phase, and the deconfined spinon state III is further confirmed by the spin excitation spectra. In the next section, we calculate and present the spin excitation spectra of the spin-1/2 tetramer chain with antiferromagnetic and ferromagnetic intra-tetramer interaction $J_2$. While the inter-tetramer coupling strength is restricted in the region $\beta\in\left(0,1\right]$. In the next section, we calculate the spin excitation spectra of the tetramer chain.

\begin{figure*}[t]
\centering
\centerline{\includegraphics[width=17.8cm]{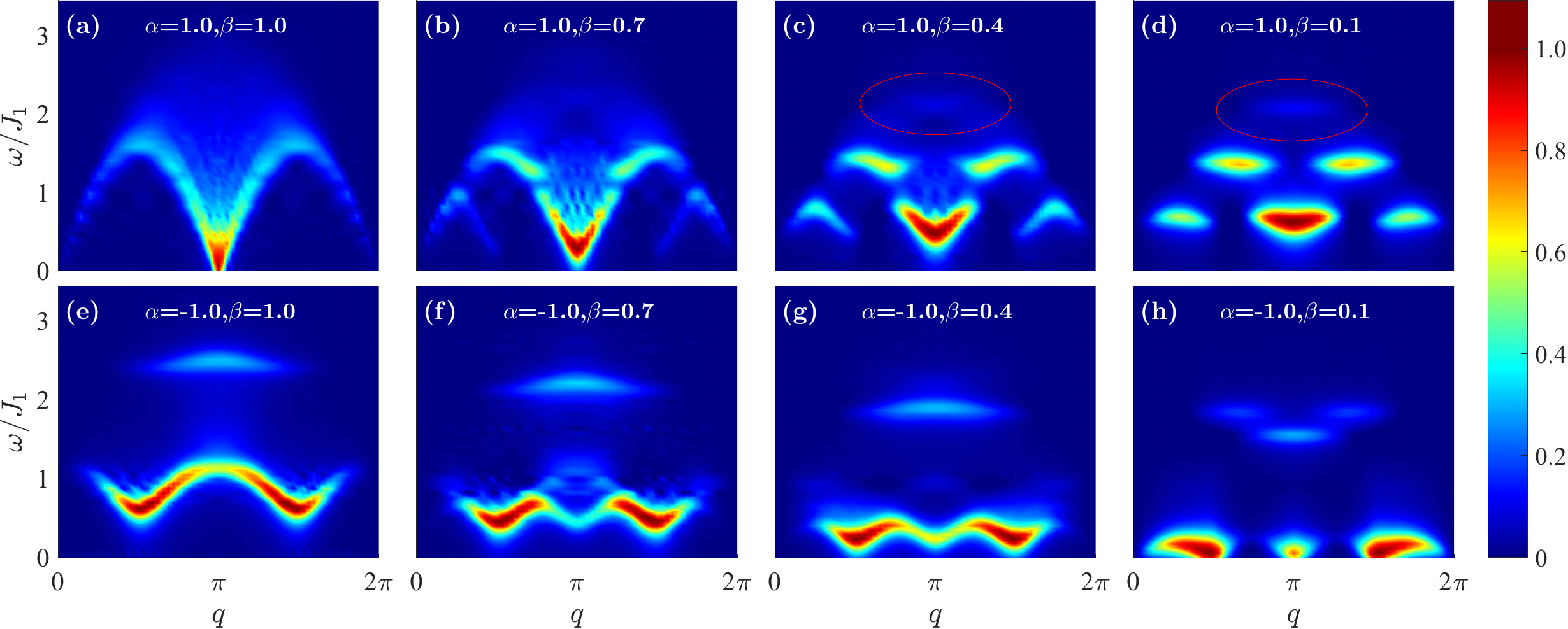}}
\caption{Dynamical structure factor $S^{zz}(q,\omega)$ calculated using Eq.~\ref{eq:Sqo} for the spin-1/2 tetramer chain with antiferromagnetic and ferromagnetic $J_2$. (a)-(d) $\alpha=1.0$. (e)-(h) $\alpha=-1.0$. The faint high-energy excitation signals are marked by red circles. $S^{zz}(q,\omega)$ are plotted using a piecewise function with the boundary value $U_0=1$. The low-intensity signals below this boundary are linear mappings of the spectral function to the color bar. The signals of $S^{zz}(q,\omega)$ above the boundary is plotted using $U=U_0+\text{log}_{10}\left[S^{zz}(q,\omega)\right]-\text{log}_{10}U_0$~\cite{ChengNPJQM2022,ChengNPJQM2024}.}
\label{fig:fig6}
\end{figure*}

\subsection{Spin excitation spectra}\label{sec:Spin excitation spectra}

In this section, we describe the spin excitation spectra of the spin-1/2 tetramer chain with different inter-tetramer exchange coupling by calculating the spin DSF using Eq.~\ref{eq:Sqo}. The intra-tetramer coupling is set to $\left|\alpha\right|=1.0$. This corresponds to the Heisenberg antiferromagnetic chain. Spin excitation spectra for $\alpha=1.0$ are shown in Figs.~\ref{fig:fig6}(a)-(d). In Fig.~\ref{fig:fig6}(a), a standard two-spinon continuum shows up when $\alpha=\beta=1.0$. The upper boundary and the lower boundary of the two-spinon continuum are described by $\omega_U\left(q\right)=\pi J_1\left|\mathrm{sin}\left(\frac{q}{2}\right)\right|$ and $\omega_L\left(q\right)=\frac{\pi}{2} J_1\left|\mathrm{sin}\left(q\right)\right|$. The two-spinon continuum achieves the highest energy $\pi J_1$ at $q=\pi$. Notably, both boundaries converge at $q=0$ and $q=2\pi$, while the spectral weight exhibits its highest intensity at the gapless point $q=\pi$.

In Fig.~\ref{fig:fig6}(b), the value of $\alpha$ is kept the same, but $\beta$ (inter-tetramer coupling) is reduced from 1.0 to 0.7. From the spin excitation spectrum we observe that the upper boundary of the two-spinon continuum undergoes a notable shift towards the lower energy levels. Concurrently, the lower boundary of the two-spinon continuum exhibits a discontinuity at the energy level $\omega=J_1$ and generates a higher and a lower energy region. The higher energy continuum displays a relatively weaker dispersion compared to its lower energy continuum. Within the lower energy continuum, the excitation spectrum is characterized by the presence of energy gaps at $q=0$, $q=\pi$, and $q=2\pi$, which indicates the existence of non-degenerate excited states. Specifically, continuous dispersion is observed in the momentum ranges of $q\in\left[\pi/4,\pi/2\right]$ and $q\in\left[3\pi/2,7\pi/4\right]$. These intervals are seamlessly connected to the remaining parts of the lower energy continuum at the boundaries $q\in\left[0,\pi/4\right]$ and $q\in\left[7\pi/4,2\pi\right]$, respectively.

As $\beta$ is decreased further, the energy dispersion breaks apart (see Fig.~\ref{fig:fig6}(c)). A distinct high-energy continuum is located at 
$\omega=2.07J_1$. This observation suggests that this high-energy continuum arises from triplon excitations with transition from the ground state $\ket{0}$ to the fourth excited state $\ket{4}$. The reason why the spectral weight of this highest energy excitation is lower than other excitations is because the higher the energy level is for a high energy excitation, the lower the transition rate (see Ref.~\cite{LiPhysRevB.111.024404}). Furthermore, both the intermediate and lowest energy continua display dispersion compared to those depicted in Fig.~\ref{fig:fig6}(b). This outcome can be attributed to two factors. Firstly, the energy gaps become larger since the lowest energy point for the lower energy continuum rises into a higher energy level. Secondly, the energy difference between the higher energy continuum and the lower energy continuum increases, which is caused by the flattening of all the energy continua while the energy levels of them remain unchanged. Regarding the origins of these continua, the intermediate energy continuum is attributed to triplon excitations with transition from the ground state $\ket{0}$ to the second excited state $\ket{2}$. Conversely, the lowest energy continuum is the result of triplon excitations with transition from the ground state $\ket{0}$ to the first excited state $\ket{1}$.

In Fig.~\ref{fig:fig6}(d), the energy continuum exhibits reduced dispersion and localization across three distinct energy levels. Specifically, the highest energy continuum is positioned at the energy level $\omega=2.07J_1$ and corresponds to the highest energy triplon excitation, transitioning from the ground state $\ket{0}$ to the excited state $\ket{4}$. The intermediate energy continuum, located at $\omega=1.36J_1$, arises from the intermediate energy triplon excitation, transitioning from $\ket{0}$ to $\ket{2}$. Meanwhile, the lowest energy continuum, situated at $\omega=0.66J_1$, pertains to the lowest energy triplon excitation, transitioning from the ground state $\ket{0}$ to the first excited state $\ket{1}$.

Spin excitation spectra for $\alpha=-1.0$ are shown in Figs.~\ref{fig:fig6}(e)-(h). In Fig.~\ref{fig:fig6}(e), we observe two energy continua. One energy continuum is at the energy level around $\omega=J_1$. The other energy continuum is at the energy level $\omega=2.39J_1$. The origin of the energy continuum at the lower energy level is the excitation from the dimer ground state singlet $\ket{0^{\prime}}$ to the dimer excited state $\ket{1^{\prime}}$. The energy continuum at the higher energy level belongs to a mixture of the excitation from tetramer singlets to triplet excited states, including $\ket{2}$ and $\ket{4}$. The lower energy continuum tends to condense the spectral weight at around $q=\pi/2$ and $q=3\pi/2$. While the lower energy continuum at $q=\pi$ has less spectral weight. It can be inferred from Fig.~\ref{fig:fig6}(e) that the spin-1/2 tetramer establishes dimer singlets and tetramer singlets along the chain when $\alpha=-1.0,\beta=1.0$, generating triplon excitations arising from both dimer singlets and tetramer singlets. 

In Fig.~\ref{fig:fig6}(f), as the inter-tetramer coupling $\beta$ is decreased from 1.0 to 0.7, the lower and the higher energy continuum are downshifted. The shape of the higher energy continuum remains unchanged. At around $q=\pi$, the spectral weight of the lower energy continuum shrinks, leaving weak spectral weight at the energy level around $\omega=J_1$. A new energy continuum with the lowest energy point at $q=\pi$ is created and is connected to the lower energy continuum at $q=3\pi/4$ and $q=5\pi/4$. For the lower energy continuum, the spectral weights still condense at $q=\pi/2$ and $q=3\pi/2$. While the spectral weight for the new energy continuum is relatively weaker. This reveals that the tetramer system for $\alpha=1.0$ and $\beta=0.7$ is dominated by dimer singlets, following by the existence of a few doublets.

In Fig.~\ref{fig:fig6}(g), both the lower and the higher energy continua are decreased into even lower energy levels when the inter-tetramer coupling $\beta$ decreased from 0.7 to 0.4. The shapes of both lower and higher energy continuum are not changed. Compared to Fig.~\ref{fig:fig6}(f), the low-energy structure observed between $q\in\left[3\pi/2,5\pi/2\right]$ is suppressed. This occurs because a reduction in the $\beta$ coupling drives the tetramer system closer to a trivial phase where all the tetramer units are isolated. The increasing number of doublets diminishes the energy gap from $\omega=0.53J_1$ to $\omega=0.29J_1$. The lower energy continuum in Fig.~\ref{fig:fig6}(g) moves to a lower energy level and becomes gapless. The gapless lower energy continuum arises from the spin-flips of the free spins and the effective spins of three-site doublets. In Fig.~\ref{fig:fig6}(h), when $\beta=0.1$, the high-energy continuum is reshaped as the tetramer system enters a phase that is a mixture of all possible combinations of tetramer singlets and three-site doublets. According to the energy analysis in Fig.~\ref{fig:fig1}, it is known that the higher energy continuum is contributed by the excitation transitions from the tetramer ground state $\ket{0}$ to the excited triplet states $\ket{2}$ and $\ket{4}$. Further detailed analysis of this low-energy excitation continuum is given in the high-energy excitations and gapless modes section. In the next section, we calculate the spin excitation spectra with a few sets of coupling strength parameters and compare them to the energy dispersion functions derived by the quantum renormalization group and perturbation theory.

\begin{figure}[t]
\centering
\centerline{\includegraphics[width=8.0cm]{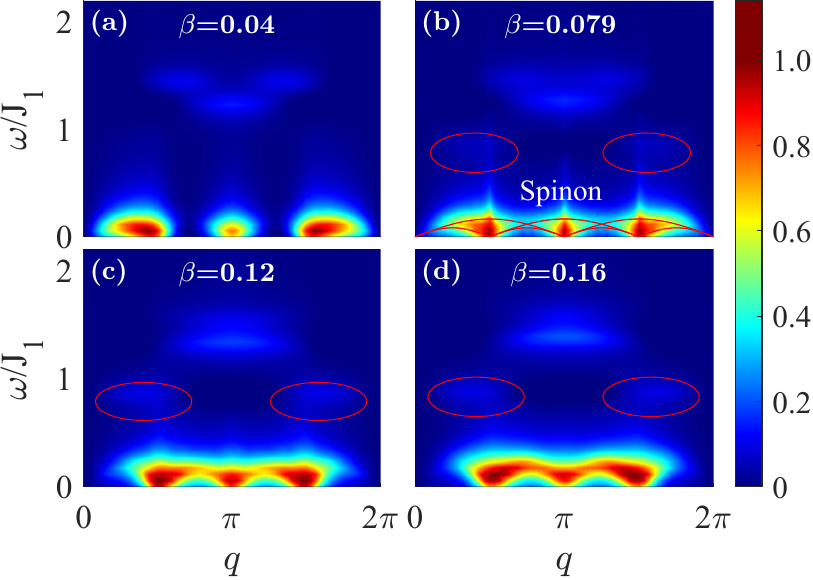}}
\caption{Dynamical structure factor with intra-tetramer coupling $\alpha=-142/240$ and inter-tetramer coupling (a) $\beta=0.04$. (b) $\beta=0.079\approx \beta_c$. The two-spinon continua surrounded by red curves are described by Eq.~\ref{eq:spinon}. (c) $\beta=0.12$. (d) $\beta=0.16$. The red circles mark the quinton excitation. The DSF is plotted using a piecewise function (stated in the caption of Fig.~\ref{fig:fig6}) with the boundary value $U_0=1$.}
\label{fig:fig7}
\end{figure}

We also calculate the spin excitation spectra using Eq.~\ref{eq:Sqo} with the parameter sets across the phase transition boundary. The spin excitation spectra results are shown in Fig.~\ref{fig:fig7}. In Fig.~\ref{fig:fig7}, we keep the intra-tetramer coupling for the tetramer system to be $\alpha=-142/240$, which is the intra-tetramer coupling parameter for CuInVO$_5$~\cite{RejaPhysRevB.99.134420}. The result in Fig.~\ref{fig:fig7}(a) is the spin excitation spectra in the tetramer phase, and the result in Fig.~\ref{fig:fig7}(b) is the spin excitation spectra right at the phase transition boundary between the tetramer phase and the Haldane phase. While the results in Fig.~\ref{fig:fig7}(c) and Fig.~\ref{fig:fig7}(d) are the spin excitation spectra in the Haldane phase. In Fig.~\ref{fig:fig7}(a), we observe low energy continuum at zero energy points and high energy continuum at energy levels $\omega=1.23J_1$ and $1.46J_1$. All the excitation signals belong to triplon excitations. In Fig.~\ref{fig:fig7}(b), the spectral weight for the low energy continuum is more localized at the zero energy points and is contracted around momentum points $q=\pi/2,\pi$ and $3\pi/2$. The lower boundary of the high energy continuum rises into the energy level $\omega=1.28J_1$, while the higher boundary of the high energy continuum remains unchanged at the energy level $\omega=1.46J_1$. Since the parameter set is at the DQCP, the low-energy continuum originates from the spinon excitation. The high-energy continuum signals are from the triplon excitations. An intermediate energy continuum signal from quinton excitations appears at $\omega=0.88J_1$.

In Fig.~\ref{fig:fig7}(c), the low-energy continuum is shifted to a high-energy level and becomes gapped. It also behaves more dispersively compared to Fig.~\ref{fig:fig7}(b). The energy level of the intermediate-energy continuum remains at the energy level $\omega=0.88J_1$. The high-energy continuum signals in Fig.~\ref{fig:fig7}(c) melt into one major signal at the energy level $\omega=1.34J_1$. In Fig.~\ref{fig:fig7}(d), the low-energy continuum enjoys a larger gapped energy. The energy level for the high-energy continuum shifts to $\omega=1.4J_1$ while the intermediate-energy continuum remains at the same energy level compared to Fig.~\ref{fig:fig7}(c). In both Fig.~\ref{fig:fig7}(c) and Fig.~\ref{fig:fig7}(d), the low-energy continuum and the high-energy continuum are contributed by triplon excitations. While the intermediate energy continuum is from quinton excitations.

\subsection{High energy excitations and gapless modes}\label{sec:High energy excitations and gapless modes}

\begin{figure}[t]
\centering
\centerline{\includegraphics[width=8.7cm]{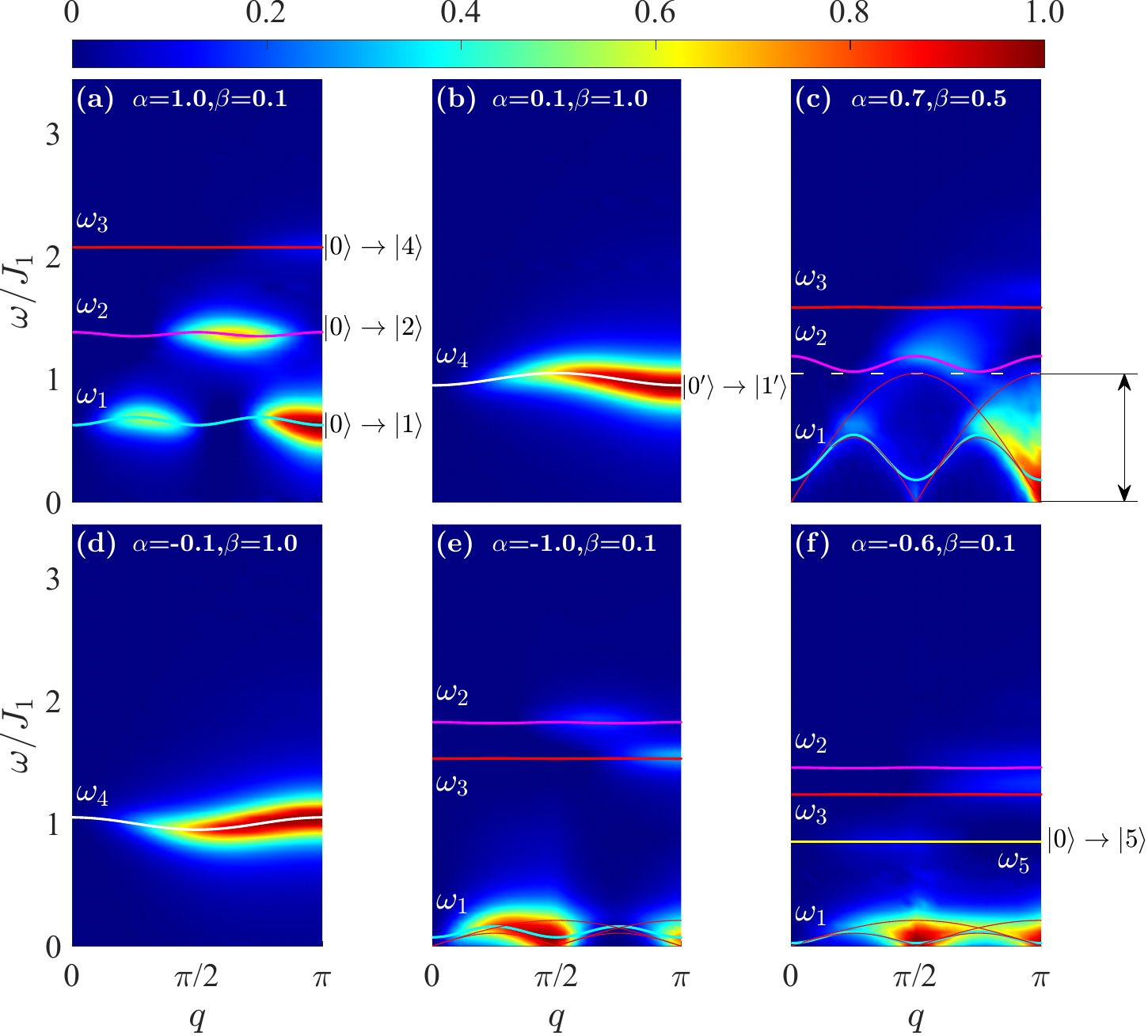}}
\caption{Dynamical structure factor for the spin-1/2 tetramer chain and the energy dispersions of spinons, triplons, and quintons. The red, blue, and white lines overlaid on the DSF intensity were computed using a quantum renormalization group approach~\cite{KargarianPhysRevA.77.032346} and perturbation theory~\cite{ChengNPJQM2022,LiPhysRevB.111.024404}. Panels (a)-(c) were calculated for $J_2>0$. Panels (d)-(f) were computed for $J_2<0$. (a) Triplon excitations in the tetramer phase. The dispersions $\omega_1, \omega_2$, and $\omega_3$ for the excited states (see Eq.~\ref{eq:e123}) and the associated state transitions are identified in the DSF plot. (b) Triplon excitations in the Haldane phase. The state transition for the $\omega_4$ branch is labeled.  The red lines in panel (c), drawn using Eq.~\ref{eq:spinon}, represent the two-spinon continuum spanning over a region marked by the double-headed arrow. The triplon excitations, identified in panel (a), are also present in this deconfined spinon state. (d) The $\omega_4$ dispersion branch of the triplon excitations in the Haldane phase. Panels (e) and (f) show the two-spinon continuum and the triplon excitations in the deconfined spinon state. The dispersion branch $\omega_5$ (see Eq.~\ref{eq:e5}) and the state transition of the quinton excitation are shown in panel (f). As before, the DSF are plotted using a piecewise function (stated in the caption of Fig.~\ref{fig:fig6}) with the boundary value $U_0=1$.}
\label{fig:fig8}
\end{figure}

\begin{figure*}[t]
\centering
\centerline{\includegraphics[width=18.0cm]{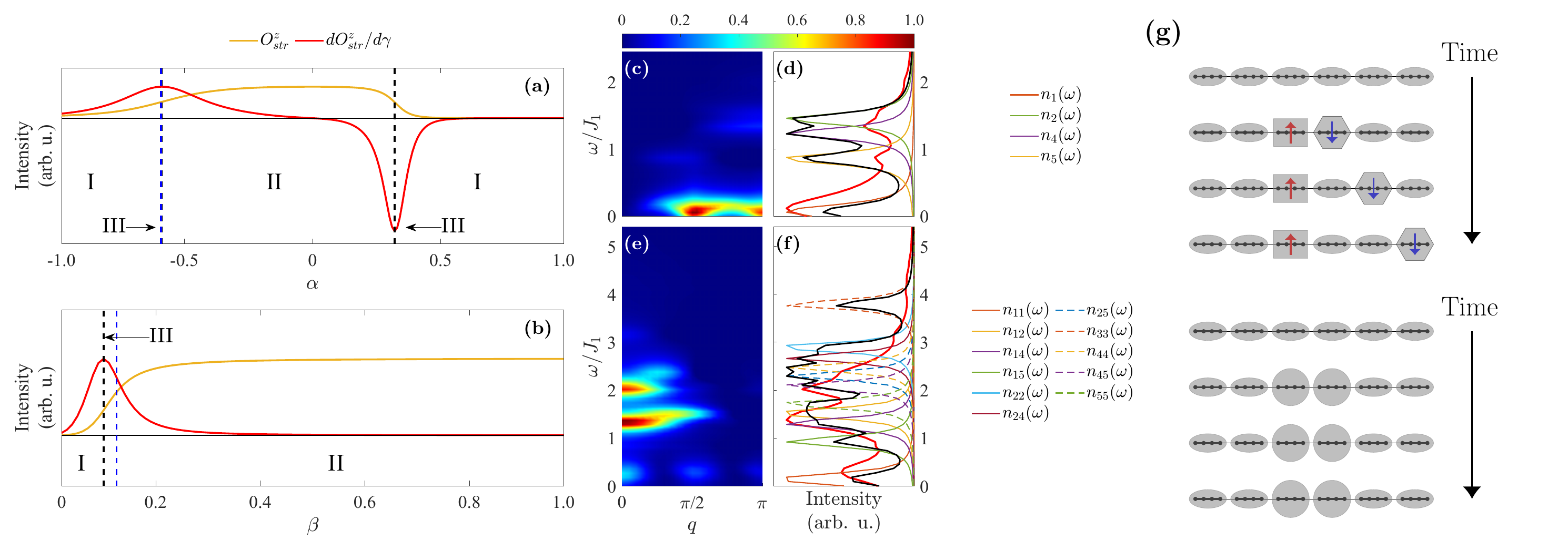}}
\caption{String order parameter, RIXS spectra, and a schematic picture of two-particle excitations for CuInVO$_5$ at the $K$-edge.  Panels (a) and (b): String order parameter and derivative. The yellow curves represent the string order parameter. The red curves are the first derivatives of the string order parameter, respectively. The Roman numerals I, II, and III are the regions for the tetramer phase, the Haldane phase, and the gapless critical phase with deconfined spinons divided by black dashed lines. The $\gamma$ in the figure legend represents $\alpha$ or $\beta$. Results in (a) and (b) are obtained by $\beta=30/240$ and $\alpha=-142/240$, respectively. Panels (c) and (e): Direct and indirect RIXS spectra. Panels (d) and (f): DOS and the integrated RIXS spectra over the momentum $q$. Black curves are the DOS spectra, while red curves are the integrated RIXS spectra $\int^{2\pi}_0S\left(q,\omega\right)dq$ in (d) or $\int^{2\pi}_0O\left(q,\omega\right)dq$ in (f). Panel (g): The upper panel is the triplon-quinton excitation. The lower panel is the two-singlon excitation, discussed in the \emph{Ground state and excited states for CuInVO$_5$} section.}
\label{fig:fig9}
\end{figure*}

In this section, we compute the DSF (described by Eq.~\ref{eq:Sqo}) to obtain the spin excitation spectra for a selected choice of parameter sets in the tetramer phase, the Haldane phase, and the deconfined spinon state. The results are shown in Fig.~\ref{fig:fig8}. We plot the high and low energy excitation dispersions overlaid on the excitation spectra in Fig.~\ref{fig:fig8}. In Fig.~\ref{fig:fig8}(a), three kinds of spin excitations can be observed when $\alpha=1.0,\beta=0.1$ and the system is in the tetramer phase, where all spins are included in tetramer singlets. The lowest, the intermediate, and the highest energy continua are at $\omega_a=E_1-E_0=0.66J_1$, $\omega_b=E_2-E_0=1.37J_1$, and $\omega_c=E_4-E_0=2.07J_1$. The origins of these excitation continuum, which are explained in the \emph{Spin excitation spectra} section, are the spin excitations from a tetramer singlet to the excited states $\ket{1}$, $\ket{2}$, and $\ket{4}$, respectively. The energy dispersion relations for these excited states are given by\begin{align}
\label{eq:e123}
\left\{
\begin{array}{ll}
\omega_1\left(q\right)&=E_1\left(J_1,J_2\right)-E_0\left(J_1,J_2\right)+A_1\mathrm{cos}\left(4q\right), \\
\; \\
\omega_2\left(q\right)&=E_2\left(J_1,J_2\right)-E_0\left(J_1,J_2\right)+A_2\mathrm{cos}\left(4q\right), \\
\; \\
\omega_3\left(q\right)&=E_4\left(J_1,J_2\right)-E_0\left(J_1,J_2\right)+A_3\mathrm{cos}\left(4q\right).
\end{array}
\right.
\end{align}

In Fig.~\ref{fig:fig8}(b), the system is in a Haldane phase according to Fig.~\ref{fig:fig3}(a), and the ground state is governed by the dimer singlet. The only existing energy continuum in the excitation spectra belongs to the triplon excitation excited from the dimer singlet $\ket{0^{\prime}}$ to the excited state $\ket{1^{\prime}}$. This energy continuum is at the energy level $\omega=E^{\prime}_1-E^{\prime}_0$, and the energy dispersion is described by the branch with $A_4$ in Eq.~\ref{eq:e4} because of the antiferromagnetic intra-tetramer interaction.

In Fig.~\ref{fig:fig8}(c), the system is in the spin-liquid state, where three-site doublets are established in the tetramer chain. The spin excitation in the spin-liquid state induces spin-flip transportation and domain wall propagation as shown in Fig.~\ref{fig:fig5}(f), resulting in a gapless excitation continuum in the excitation spectra. Using quantum renormalization group analysis (see Sec.~B in the Supplementary Note) of the spin-1/2 tetramer chain, we can compute the lower and the higher boundaries of the gapless low-energy two-spinon continuum, which is described by\begin{align}
\label{eq:spinon}
\left\{
\begin{array}{ll}
\omega^{\prime}_L\left(q\right)&=\frac{\pi}{2}J_{eff}\left|\mathrm{sin}\left(2q\right)\right|, \\
\; \\
\omega^{\prime}_U\left(q\right)&=\pi J_{eff}\left|\mathrm{sin}\left(q\right)\right|, \\
\; \\
\omega^{\prime\prime}_U\left(q\right)&=\pi J_{eff}\left|\mathrm{sin}\left(q-\frac{\pi}{2}\right)\right|,
\end{array}
\right.
\end{align} where $J_{eff}=2\pi\beta/3$ is the effective exchange interaction. In the low-energy two-spinon continuum, we can see gapless points at $q=\pi/2$ and $q=\pi$, which belong to the intra-tetramer spinon propagation and the inter-tetramer spinon propagation, respectively. The tetramer system shows a stronger spectral weight at the zero energy point around $q=\pi$ compared to other regions, which indicates that the spinons tend to propagate within a tetramer instead of propagating between tetramers.  The energy dispersion relations derived by the perturbation theory (see Sec.~C in the Supplementary Note) can also describe some regions of the gapless low-energy excitation continuum. The lowest energy dispersion $\omega_1\left(q\right)$ describes the lower boundary of the gapless two-spinon continuum at the energy level around $\omega=0.36J_1$, but never touches the zero energy points at the gapless momentum point $q=0$, $q=\pi/2$, and $q=\pi$. The intermediate energy dispersion $\omega_2\left(q\right)$ describes the upper boundary of the gapless low-energy excitation continuum at the energy level $\omega=1.12J_1$. The highest energy dispersion $\omega_3\left(q\right)$, referring to the triplon excitation, describes the highest energy continuum around its highest energy point. Note, the renormalization analysis and the perturbation theory approaches are complementary to each other.

In Fig.~\ref{fig:fig8}(d), the system is in a Haldane phase, which is close to the situation in Fig.~\ref{fig:fig8}(b). The energy continuum is at the energy level of $\omega=E^{\prime}_1-E^{\prime}_0$, which is the same energy level of the energy continuum in Fig.~\ref{fig:fig8}(b). Note, the energy dispersion amplitude is inversed to the dispersion amplitude in $A_4$ due to the intra-tetramer coupling changing from antiferromagnetic to ferromagnetic. Therefore, the energy dispersion branch, labeled as $-A_4$, in Fig.~\ref{fig:fig8}(d) is described by\begin{equation}
\label{eq:e4}
\omega_4\left(q\right)=E^{\prime}_1-E^{\prime}_0\pm A_4\mathrm{cos}\left(4q\right).
\end{equation} In Fig.~\ref{fig:fig8}(e), we observe both the high-energy continuum and the gapless low-energy continuum since the system is in the tetramer phase. The high-energy continuum includes two kinds of high-energy excitations. The energy curve $\omega_3(q)$ is the excitation from the tetramer ground state singlet $\ket{0}$ to the excited state $\ket{3}$, and the energy curve $\omega_2(q)$ refers to the excitation from the ground state $\ket{0}$ to the excited state $\ket{2}$. The upper and the lower boundary of low-energy continuum with gapless modes at $q=\pi/4$ and $q=\pi/2$ in Fig.~\ref{fig:fig8}(e) are also described by Eq.~\ref{eq:spinon}. In Fig.~\ref{fig:fig8}(f), we still observe the gapless low-energy excitation because that system is still in when $\alpha=-0.6$ and $\beta=0.1$. Due to the diminished ferromagnetic intra-tetramer interaction coupling $\alpha$, the energy levels for the high energy excitations are depressed, and the differences of the energy levels for high energy dispersions $\omega_2(q)$ and $\omega_4(q)$ shrink compared to Fig~\ref{fig:fig8}(e). An energy continuum appears at the energy level below $\omega_4(q)$, which is contributed by the quinton excitation (excitation from $\ket{0}$ to $\ket{5}$) according to the perturbation theory and is described by the energy relation curve $\omega_5(q)$, which is described by the following equation\begin{equation}
\label{eq:e5}
\omega_5\left(q\right)=E_5\left(J_1,J_2\right)-E_0\left(J_1,J_2\right)+A_5\mathrm{cos}\left(4q\right).
\end{equation}
The energy dispersion relation presented the quinton excitation in Eq.~\ref{eq:e5}, which indicates that the excitation from $\ket{0}$ to $\ket{5}$ is highly localized compared to other high-energy excitations.

\subsection{Ground state and excited states for CuInVO$_5$}\label{sec:Ground state and excited states for CuInVO$_5$}

In this section, we study the ground state of CuInVO$_5$, which is a candidate Haldane-like material that can be realized in a 1D tetramer spin chain~\cite{RejaPhysRevB.99.134420}. We begin by calculating the string order parameter. Next, the direct and the indirect RIXS spectra are computed to investigate the spin dynamics of the spin-1/2 tetramer chain. In these calculations, the intra-tetramer and inter-tetramer interactions are $\alpha=-142/240$ and $\beta=30/240$, which belong to the exchange couplings of CuInVO$_5$~\cite{RejaPhysRevB.99.134420}. The string order parameter and its derivatives are shown in Figs.~\ref{fig:fig9}(a)-(b). In Fig.~\ref{fig:fig9}(a), we show the corresponding string order parameter with inter-tetramer coupling $\beta=30/240$ and intra-tetramer coupling $\alpha$ in the range $\alpha\in\left[-1.0,1.0\right]$. The material parameters of the spin-1/2 tetramer chain place it in the tetramer phase when $\alpha=-1.0$. As $\alpha$ is increased from -1.0 to 1.0, the string order parameter rises and becomes a plateau around $\alpha=-0.4$ and decreases at $\alpha=0.3$ (see Fig.~\ref{fig:fig9}(a)). During this transition, the ground state of the spin chain transitions between a tetramer phase, the Haldane phase, and the intermediate deconfined quantum critical state. The first-order derivative reaches its maximum values for $\alpha<0$ and decreases to the minimum value for $\alpha>0$. In Fig.~\ref{fig:fig9}(b), we show the corresponding string order parameter with intra-tetramer coupling $\alpha=-142/240$ and inter-tetramer coupling $\beta$ in the range $\beta\in\left[0,1.0\right]$. As $\beta$ increases from 0, the string order parameter increases into a platform at about $\beta=0.2$. The ground state passes through the tetramer phase and the Haldane phase. The first-order derivative $dO^z_{str}/d\beta$ obtains its maximum value in the deconfined spinon state.

In Fig.~\ref{fig:fig9}(c) and Fig.~\ref{fig:fig9}(e), we present the direct ($L$-edge) and indirect ($K$-edge) RIXS spectra in the momentum region $q\in[0,\pi]$. We compare the integrated RIXS spectra and the density of states (DOS) spectra in Fig.~\ref{fig:fig9}(d) and Fig.~\ref{fig:fig9}(f). In Fig.~\ref{fig:fig9}(c), three energy continua can be observed in the direct RIXS spectra. For the lowest energy continuum at $\omega=0.12J_1$, the spectral weights are condensed at $q=\pi/2$ and $q=\pi$. This lowest energy continuum contributed by the triplon excitation with the excited state $\ket{5}$ is gapped, and no gapless excitations are included, which can be further confirmed from the lowest energy continuum in Fig.~\ref{fig:fig9}(e). The intermediate energy continuum at the energy level $\omega=0.88J_1$ belongs to the quinton excitation. The highest energy continuum at the energy level $\omega=1.34J_1$ is contributed by triplon excitations with both excited states $\ket{2}$ and $\ket{4}$. In Fig.~\ref{fig:fig9}(d), three major direct RIXS spectra signals can be seen at the energy levels $\omega=0.12J_1$, $\omega=0.88J_1$, and $\omega=1.34J_1$. The energy levels for the DOS signals are $\omega=0.06J_1$, $\omega=0.88J_1$, $\omega=1.23J_1$, and $\omega=1.46J_1$, indicating that the energy levels for the direct RIXS spectra and the DOS signals are consistent. 

In Fig.~\ref{fig:fig9}(e), we present the indirect RIXS spectra. The spectral weights of the lowest energy continuum are strong at $q=0$, $q=\pi/2$, and $q=\pi$. The gapped lowest energy continuum indicates that the ground state of the CuInVO$_5$ generates gapped excitation and the system is in the Haldane phase. The existence of this gap, although not very clear in the $L$-edge spectrum (Fig.~\ref{fig:fig9}(c)), is clearly revealed in the indirect $K$-edge RIXS spectrum. At the zero momentum point, higher energy continuum at the energy levels around $\omega=1.38J_1$ and $\omega=2.02J_1$ are localized. An additional RIXS signal appears at the energy level $\omega=2.3J_1$ and around the momentum $q=\pi/4$.

The integrated indirect RIXS spectra and the corresponding DOS spectra are compared in Fig.~\ref{fig:fig9}(f). The energy levels for the integrated indirect RIXS signals are $\omega=0.28J_1$, $\omega=1.38J_1$, $\omega=2.02J_1$, $\omega=3.12J_1$, and $\omega=4.03J_1$. According to the DOS spectra, the integrated RIXS signal at the energy level $\omega=0.28J_1$ arises from the double-triplon excitation with the excited state $\ket{1}$. At the energy level $\omega=1.38J_1$, the integrated RIXS signal contains two-triplon excitation, the triplon-quinton excitation, and the two-quinton excitation.  At the energy level $\omega=2.02J_1$, the integrated RIXS signal is from the two-triplon and the triplon-quinton excitations. However, the signals for the two-particle excitations are hard to be detected due to the lower transition probabilities in the higher energy region. There are two integrated RIXS signals at energy levels higher than $\omega=2.02J_1$, which are at $\omega=3.12J_1$ and $\omega=4.03J_1$. The highest energy integrated RIXS signal is from the two-singlon excitation, whose DOS signal is at the energy level $\omega=3.76J_1$. The second highest integrated RIXS signal refers to the two-triplon excitation, which corresponds to the DOS signal at the energy level $\omega=2.93J_1$. The integrated RIXS spectra and DOS signals show that the RIXS spectra are good at capturing the two-particle excitations below the energy level $\omega=3.12J_1$. But the DOS spectra at the energy level higher than $\omega=3.12J_1$ experiences a downshift compared to the integrated RIXS spectra, which is the result of the scale effect of the perturbation theory and the low transition probabilities for high-energy excitations. The $K$-edge RIXS creates two-particle excitations, including two-triplon, triplon-quinton, and two-singlon excitations. In Fig.~\ref{fig:fig9}(g), we give a schematic picture for the triplon-quinton excitation (upper panel) and two-singlon excitation (lower panel) to show the formation and time evolution of the two-particle excitations.

In addition, we calculate the transition rates for the single-spin and two-particle excitations. Sec.~D of the Supplementary Note lists the equations utilized to calculate these rates. Tables I-II (in the Supplementary Note) summarize the transition rates between the various quantum states of the tetramer chain. In Table I, the transition rate of the triplon excitation transition process ($\ket{0}\rightarrow\ket{1}$) in both the single tetramer unit and the 2-tetramer system has an enhanced value compared to the transition rates from the transition processes $\ket{0}\rightarrow\ket{2}$ and $\ket{0}\rightarrow\ket{4}$. This result is consistent with the $L$-edge RIXS spectra in Fig.~\ref{fig:fig9}(c), where the triplon excitation channel $\ket{0}\rightarrow\ket{1}$ gives rise to the strongest signal, while the signals from the channels $\ket{0}\rightarrow\ket{2}$ and $\ket{0}\rightarrow\ket{4}$ are relatively suppressed. The transition rate for the quinton excitation transition process $\ket{0}\rightarrow\ket{5}$ is zero in a single tetramer unit. However, the transition rate process $\ket{0}\rightarrow\ket{5}$ evolves into a non-zero value when considering a system with two coupled tetramer units. This indicates that the quinton excitation is influenced by the non-local effect of the inter-tetramer interaction. Thus, the coupling between tetramers assists in the creation of the quinton excitation.

We display the transition rates for two-particle excitations in Table II. We find that the transition rates of two-particle excitations, which include the quinton, are all zero. This characteristic reveals that even in the two-particle excitation process, the excitations, including the quinton, are non-local excitations and only survive in the coupled tetramer system. Hence, we only focus on the transition rates for the coupled tetramer unit. Based on the transition rates shown in Table II for the coupled tetramer system, we can conclude that the strongest excitation signal is at the energy level $\omega=1.38J_1$. This signal arises from $\ket{0}\ket{0}\rightarrow\ket{1}\ket{2}$, $\ket{0}\ket{0}\rightarrow\ket{1}\ket{4}$, and $\ket{0}\ket{0}\rightarrow\ket{1}\ket{5}$, which have the strongest total transition rate (3.729). The integrated RIXS spectra signal at the lowest energy level $\omega=0.28J_1$ is contributed by $\ket{0}\ket{0}\rightarrow\ket{1}\ket{1}$, with the second highest transition rate (2.486). The integrated RIXS spectra signals at the lowest energy level $\omega=2.02J_1$ containing transitions $\ket{0}\ket{0}\rightarrow\ket{2}\ket{5}$, $\ket{0}\ket{0}\rightarrow\ket{4}\ket{4}$, and $\ket{0}\ket{0}\rightarrow\ket{4}\ket{5}$ have the lowest transition rate (4.625$\times 10^{-4}$). In contrast, the excitation signal in the RIXS spectra at the lowest energy level $\omega=0.28J_1$ is weaker than the RIXS spectra signals at the energy level $\omega=2.02J_1$. The transition rate for $\ket{0}\ket{0}\rightarrow\ket{3}\ket{3}$ is zero. However, non-zero signals at the energy level of $\omega=4J_1$ are observable in Fig.~\ref{fig:fig8}(f). This implies that the perturbation theory method is more powerful when predicting the signal strength in single-spin excitation compared to the two-particle features. All the RIXS and transition rate calculations suggest that the quinton excitation exists in the spin-1/2 tetramer chain.

\section{Discussion}\label{sec:Discussion}

We have examined the physics of a spin-1/2 tetramer chain, which consists of repeated units of four coupled spins, forming tetramers along a 1D lattice. Based on the relative intra-tetramer and inter-tetramer competing exchange interaction strengths, the system can transition between a tetramer phase~\cite{RejaPhysRevB.99.134420}, a Haldane-like phase~\cite{Haldane1983464,WenPhysRevB.80.155131,FidkowskiPhysRevB.83.075103,ShapourianPhysRevLett.118.216402,GongPhysRevB.78.104416,AffleckPhysRevLett.59.799}, and an intermediate deconfined quantum critical state~\cite{Senthilscience.1091806,JiangPhysRevB.99.075103,RobertsPhysRevB.99.165143,HuangPhysRevB.100.125137,Pronks53y-qmr4,YangPhysRevE.104.064121,LiPhysRevB.107.085130}. The tetramer phase is gapped trivial but can support exotic triplon or quinton excitations. The ground state of the system transforms from a gapless critical deconfined spinon state to the gapped Haldane phase, which is an example of a quantum phase transition that is described by the concept of deconfined quantum criticality. Conceptually, the spinons exist as free gapless modes in the deconfined spinon state (which we labeled as III). However, with increasing bond dimerization interaction, an instability of the deconfined spinons causes them to form bound pairs. This confinement leads to the generation of a mass gap in the excitation spectrum and eventually leads to the onset of the Haldane-like phase. The Haldane-like phase is a SPT phase whose existence is revealed by a non-vanishing string order parameter that is capable of detecting a broken hidden $Z_2\times Z_2$ discrete symmetry. We show that in the spin-1/2 tetramer chain, the DSF (captured in the INS and the $L$-edge RIXS spectrum) is sensitive to a wide variety of excitations ranging from low to high energy, gapless to gapped, and trivial to exotic. Our calculations demonstrate that x-ray spectroscopy has the ability to comprehensively detect spin order and spin excitations of a candidate Haldane-like material (CuInVO$_5$), which can be described by a spin-1/2 tetramer chain.

Finally, note that the instrumentation resolutions of $L$-edge and $K$-edge RIXS setups are approximately 15 meV and 25 meV, respectively. The lowest and highest energy signals for the calculated $L$-edge RIXS spectrum are at the energy levels of 3.5 meV and 27.7 meV. The lowest and highest energy signal, for the calculated $K$-edge RIXS spectrum (which includes the two-particle features) are at the energy levels of 5.7 meV and 41.4 meV. Therefore, at least the high-energy signals for both the $L$-edge and $K$-edge RIXS results predicted in our computations for CuInVO$_5$ can be experimentally detected. Currently, there is no RIXS or INS data for CuInVO$_5$. Hopefully, our predictions of novel excitations in the spin-1/2 tetramer chain will inspire the experimental community to explore this compound.


\section{Method}\label{sec:Method}

\subsection{Density matrix renormalization group}
In this section, we discuss the numerical method for calculating the string order operator, the single-spin excitation spectra, the RIXS spectra, and the DOS. The string order parameter $O^z_{str}\left(\alpha,\beta\right)$ is calculated using the DMRG algorithm. Note, the string operator $\Theta_{4n,4n^{\prime}+1}$ is the expectation of a matrix product operator, which is expressed as\begin{equation}
\label{eq:product}
\begin{split}
\Theta_{4n,4n^{\prime}+1}=-&\left\langle S^z_{4n}e^{i\pi\left(S^z_{4n+1}+S^z_{4n+2}+\cdot\cdot\cdot+S^z_{4n^{\prime}-1}+S^z_{4n^{\prime}}\right)}S^z_{4n^{\prime}+1}\right\rangle \\
=-&\langle I\otimes I\otimes I\otimes S^z\otimes\sigma^z\otimes\cdot\cdot\cdot\otimes\sigma^z\otimes S^z\otimes I\otimes I\otimes I\rangle,
\end{split}
\end{equation} where $S^z$ is the spin matrix, $\sigma^z$ is the Pauli matrix, and $I$ is the $2\times 2$ identity matrix. Therefore, the string order operator is obtained by calculating the ground-state spin expectations using DMRG. The spin excitation spectrum at the $L$-edge RIXS (Eq.~\ref{eq:Sqo}) and $K$-edge RIXS (Eq.~\ref{eq:Oqo}) are computed using the Krylov-space CV method in the DMRG framework~\cite{NoceraPhysRevE.94.053308}. The Krylov-space CV method calculates the excitation spectra with high accuracy and efficiency compared to the traditional CV method~\cite{KuhnerPhysRevB.60.335,JeckelmannPhysRevB.66.045114,UmmethumPhysRevB.86.104403,NoceraPhysRevE.94.053308}. Both the direct and the indirect RIXS intensity response functions are computed within the UCL expansion scheme~\cite{Brink_2007,NagNatComm2022}. According to the selection rules, the $L$-edge RIXS spectrum in Fig.~\ref{fig:fig9}(c) includes the signals for single-spin excitations, which are processes where the singlet ground state with $\mathrm{M}_\mathrm{s}=0$ is excited into a state with $\mathrm{M}_\mathrm{s}=\pm 1$. The $K$-edge RIXS spectrum in Fig.~\ref{fig:fig9}(e) detects two-particle excitation signals contributed by the processes where two tetramers in the singlet ground state transition to excited states. In this two-tetramer case, the change of $\mathrm{M}_\mathrm{s}$ can give rise to two different cases. In one case, the excited states have $\mathrm{M}_\mathrm{s}=1$ in one tetramer and $\mathrm{M}_\mathrm{s}=-1$ in another. In another case, both the excited states have $\mathrm{M}_\mathrm{s}=0$ in the two tetramers.

The single-spin excitation spectra and the $L$-edge RIXS spectra are calculated using a two-spin correlation function\begin{equation}
\label{eq:Sqo}
S(q,\omega)=\sum_{\alpha=x,y,z}\sum_f|\bra{g}\hat{S}^{\alpha}_q\ket{f}|^2\delta(\omega+\omega_g-\omega_f),
\end{equation} where $\omega$ is the energy, $\ket{g}$ is the ground state, $\ket{f}$ is the final state, and $\omega_f-\omega_g$ is the resonant energy of the single spin-flip excitation. The single-spin form factor is $S^{\alpha}_q=\frac{1}{\sqrt{N}}\sum_{j}e^{iqr_j}S^{\alpha}_j$, where $r_j$ is the site position and $S^{\alpha}_j$ represents the spin operator on the $j$-th site. The $K$-edge RIXS spectra, which captures the double spin-flip excitation (also called the two-particle excitation), is calculated using a four-spin correlation function\begin{equation}
\label{eq:Oqo}
O(q,\omega)=\sum_f|\bra{g}O_q\ket{f'}|^2\delta(\omega+\omega_g-\omega_{f'}),
\end{equation} where $\ket{f'}$ is the final state of the $K$-edge RIXS scattering process and $\omega_{g}-\omega_{f'}$ is the resonant energy of the double spin-flip excitation. The two-spin form factor is expressed as $O_q=\frac{1}{\sqrt{N}}\sum_je^{iqr_j}\hat{\bs}_j\cdot\hat{\bs}_{j+1}$. The calculations of Eqs.~\ref{eq:Sqo} and~\ref{eq:Oqo} are carried out using the CV method and the spectral broadening is set to be 0.1$J_1$.

The DMRG calculation in this paper is performed using a lattice length $L=64$, which is a computationally adequate length as seen from the excellent agreement between the numerically computed spectra and the analytical results (both quantum renormalization group analysis and perturbation theory). For our DMRG computations, the maximum number of kept states is $m=800$ and the truncation error is set to $\varepsilon=1\times 10^{-8}$. The excitation spectra in Figs.~\ref{fig:fig6}-\ref{fig:fig8} are plotted using a piecewise function with the boundary value $U_0=1$. The excitation spectra are plotted using linear mappings for the intensity strengths lower than $U_0=1$. And the intensity strengths larger than $U_0=1$ are plotted using the function $U=U_0+\text{log}_{10}\left[S^{zz}(q,\omega)\right]-\text{log}_{10}U_0$~\cite{ChengNPJQM2022,ChengNPJQM2024}. The current calculation parameters, including the lattice length and the maximum number of kept states, ensure that the calculation results converge (see Sec.~G in the Supplementary Note).

\subsection{Density of states}

The DOS for single-spin excitations shown in Fig.~\ref{fig:fig9}(d) are calculated using\begin{equation}
\label{eq:n1}
n^{\alpha}(q,\omega)=-\frac{1}{\pi}\sum^{2\pi}_{q=0}\mathrm{Im}\left[\frac{1}{\omega+i\delta-\omega_{\alpha}(q)}\right],\alpha=1\cdots5,
\end{equation} where $\alpha$ represents the various energy levels. The DOS for two-particle excitations shown in Fig.~\ref{fig:fig9}(f) are calculated using\begin{equation}
\label{eq:n2}
n^{\alpha\beta}(q,\omega)=-\frac{1}{\pi}\sum^{2\pi}_{k,q=0}\mathrm{Im}\left[\frac{1}{\omega+i\delta-\omega_{\alpha}(k-q)-\omega_{\beta}(q)}\right].
\end{equation} The possible allowed combinations of $\alpha$ and $\beta$ are $\{\alpha,\beta\}\in\{(1,1),(1,2),(1,4),(1,5),(2,2),(2,4),(2,5),(3,3),(4,4),(4,5),$ $(5,5)\}$. The spectral broadening for the DOS formulae (Eqs.~\ref{eq:n1} and \ref{eq:n2}) is $\delta=0.1J_1$. The transition rates were computed using Eqs.~13 and 14 stated in the Supplementary Note. The transition rate results and descriptions are summarized in Tables I and II in Sec.~D of the Supplementary Note.

\section{Data availability}

The datasets generated and analyzed during the current study are not publicly available because they consist of large-scale numerical simulation data and intermediate files that are not stored in a standardized public repository, but are available from the corresponding author on reasonable request.

\section{Code availability}

The codes generated and used during the current study are not publicly available because they are part of an actively developed research codebase and are not maintained as a public repository, but are available from the corresponding author on reasonable request.

\section{acknowledgements}
J.L.L. and D.X.Y. thank Muwei Wu for helpful discussions. J.L.L. and D.X.Y. are supported by NKRDPC2022YFA1402802, NSFC-92165204, NSFC-12494591, 
 Leading Talent Program of Guangdong Special Projects (No. 201626003), Guangdong Provincial Key Laboratory of Magnetoelectric Physics and Devices (No. 2022B1212010008), Research Center for Magnetoelectric Physics of Guangdong Province (No. 2024B0303390001), and Guangdong Provincial Quantum Science Strategic Initiative (No. GDZX2401010). T.D. acknowledges Augusta University High Performance Computing Services (AUHPCS) for providing computational resources contributing to the results presented in this publication. J.Q.C. is supported by Special Project in Key Areas for Universities in Guangdong Province (No. 2023ZDZX3054) and Dongguan Key Laboratory of Artificial Intelligence Design for Advanced Materials (DKL-AIDAM).

\section{Author contributions}

D.X.Y., J.L., and T.D. conceived and designed the project. J.L. and J.Q.C. performed the DMRG simulations, the perturbation theory, and the quantum renormalization calculations. T.D. and D.X.Y checked the calculations. All authors contributed to the discussion and interpretation of the results. J.L., T.D., and D.X.Y. wrote the paper.

\section{Competing interests}

J.L., J.Q.C., T.D., and D.X.Y. declare no competing financial or non-financial Interests.

\section*{References}
\bibliography{ref}

@article{ChengNPJQM2022,
title = {Fractional and composite excitations of antiferromagnetic quantum spin trimer chains},
journal = {npj Quantum Mater.},
volume = {7},
pages = {3},
year = {2022},
issn = {2397-4648},
doi = {10.1038/s41535-021-00416-4},
url = {https://doi.org/10.1038/s41535-021-00416-4},
author = {Cheng, Jun-Qing and Li, Jun and Xiong, Zijian and Wu, Han-Qing and Sandvik, Anders W. and Yao, Dao-Xin}
}

@article{ChengNPJQM2024,
      title={Quantum phase transition and composite excitations of antiferromagnetic spin trimer chains in a magnetic field}, 
      journal = {npj Quantum Mater.},
      volume = {9},
      pages = {96},
      year={2024},
      issn = {2397-4648},
      doi = {10.1038/s41535-024-00705-8},
      url={https://doi.org/10.1038/s41535-024-00705-82},
      author={Jun-Qing Cheng and Zhi-Yao Ning and Han-Qing Wu and Dao-Xin Yao}
}

@article{LiPhysRevB.111.024404,
  title = {Resonant inelastic x-ray scattering spectra of spinon, doublon, and quarton excitations of a spin-$\frac{1}{2}$ antiferromagnetic {H}eisenberg trimer chain},
  author ={Li, Junli and Cheng, Jun-Qing and Datta, Trinanjan and Yao, Dao-Xin},
  journal = {Phys. Rev. B},
  volume = {111},
  issue = {2},
  pages = {024404},
  numpages = {13},
  year = {2025},
  month = {Jan},
  publisher = {American Physical Society},
  doi = {10.1103/PhysRevB.111.024404},
  url = {https://link.aps.org/doi/10.1103/PhysRevB.111.024404}
}

@article{SuPhysRevB.78.104416,
  title = {Magnetization plateaus, {Haldane-like} gap, string order, and hidden symmetry in a spin-$\frac{1}{2}$ tetrameric {Heisenberg} antiferromagnetic chain},
  author = {Gong, Shou-Shu and Su, Gang},
  journal = {Phys. Rev. B},
  volume = {78},
  issue = {10},
  pages = {104416},
  numpages = {10},
  year = {2008},
  month = {Sep},
  publisher = {American Physical Society},
  doi = {10.1103/PhysRevB.78.104416},
  url = {https://link.aps.org/doi/10.1103/PhysRevB.78.104416}
}

@article{WenPhysRevB.80.155131,
  title = {Tensor-entanglement-filtering renormalization approach and symmetry-protected topological order},
  author = {Gu, Zheng-Cheng and Wen, Xiao-Gang},
  journal = {Phys. Rev. B},
  volume = {80},
  issue = {15},
  pages = {155131},
  numpages = {23},
  year = {2009},
  month = {Oct},
  publisher = {American Physical Society},
  doi = {10.1103/PhysRevB.80.155131},
  url = {https://link.aps.org/doi/10.1103/PhysRevB.80.155131}
}

@article{FidkowskiPhysRevB.83.075103,
  title = {Topological phases of fermions in one dimension},
  author = {Fidkowski, Lukasz and Kitaev, Alexei},
  journal = {Phys. Rev. B},
  volume = {83},
  issue = {7},
  pages = {075103},
  numpages = {13},
  year = {2011},
  month = {Feb},
  publisher = {American Physical Society},
  doi = {10.1103/PhysRevB.83.075103},
  url = {https://link.aps.org/doi/10.1103/PhysRevB.83.075103}
}

@article{ShapourianPhysRevLett.118.216402,
  title = {Many-Body Topological Invariants for Fermionic Symmetry-Protected Topological Phases},
  author = {Shapourian, Hassan and Shiozaki, Ken and Ryu, Shinsei},
  journal = {Phys. Rev. Lett.},
  volume = {118},
  issue = {21},
  pages = {216402},
  numpages = {6},
  year = {2017},
  month = {May},
  publisher = {American Physical Society},
  doi = {10.1103/PhysRevLett.118.216402},
  url = {https://link.aps.org/doi/10.1103/PhysRevLett.118.216402}
}

@article{HanPhysRevB.96.125105,
  title = {Boundary conformal field theory and symmetry-protected topological phases in $2+1$ dimensions},
  author = {Han, Bo and Tiwari, Apoorv and Hsieh, Chang-Tse and Ryu, Shinsei},
  journal = {Phys. Rev. B},
  volume = {96},
  issue = {12},
  pages = {125105},
  numpages = {15},
  year = {2017},
  month = {Sep},
  publisher = {American Physical Society},
  doi = {10.1103/PhysRevB.96.125105},
  url = {https://link.aps.org/doi/10.1103/PhysRevB.96.125105}
}

@article{GongPhysRevB.78.104416,
  title = {Magnetization plateaus, {H}aldane-like gap, string order, and hidden symmetry in a spin-$\frac{1}{2}$ tetrameric {H}eisenberg antiferromagnetic chain},
  author = {Gong, Shou-Shu and Su, Gang},
  journal = {Phys. Rev. B},
  volume = {78},
  issue = {10},
  pages = {104416},
  numpages = {10},
  year = {2008},
  month = {Sep},
  publisher = {American Physical Society},
  doi = {10.1103/PhysRevB.78.104416},
  url = {https://link.aps.org/doi/10.1103/PhysRevB.78.104416}
}

@article{Haldane1983464,
title = {Continuum dynamics of the 1-{D} {H}eisenberg antiferromagnet: Identification with the {O}(3) nonlinear sigma model},
author = {F.D.M. Haldane},
journal = {Physics Letters A},
volume = {93},
number = {9},
pages = {464-468},
year = {1983},
issn = {0375-9601},
doi = {https://doi.org/10.1016/0375-9601(83)90631-X},
url = {https://www.sciencedirect.com/science/article/pii/037596018390631X}
}

@article{AffleckPhysRevLett.59.799,
  title = {Rigorous results on valence-bond ground states in antiferromagnets},
  author = {Affleck, Ian and Kennedy, Tom and Lieb, Elliott H. and Tasaki, Hal},
  journal = {Phys. Rev. Lett.},
  volume = {59},
  issue = {7},
  pages = {799--802},
  numpages = {0},
  year = {1987},
  month = {Aug},
  publisher = {American Physical Society},
  doi = {10.1103/PhysRevLett.59.799},
  url = {https://link.aps.org/doi/10.1103/PhysRevLett.59.799}
}

@article{ScheiePhysRevB.103.224434,
  title = {Witnessing entanglement in quantum magnets using neutron scattering},
  author = {Scheie, A. and Laurell, Pontus and Samarakoon, A. M. and Lake, B. and Nagler, S. E. and Granroth, G. E. and Okamoto, S. and Alvarez, G. and Tennant, D. A.},
  journal = {Phys. Rev. B},
  volume = {103},
  issue = {22},
  pages = {224434},
  numpages = {16},
  year = {2021},
  month = {Jun},
  publisher = {American Physical Society},
  doi = {10.1103/PhysRevB.103.224434},
  url = {https://link.aps.org/doi/10.1103/PhysRevB.103.224434}
}

@article{HakobyanPhysRevB.63.144433,
  title = {Phase diagram of the frustrated two-leg ladder model},
  author = {Hakobyan, T. and Hetherington, J. H. and Roger, M.},
  journal = {Phys. Rev. B},
  volume = {63},
  issue = {14},
  pages = {144433},
  numpages = {17},
  year = {2001},
  month = {Mar},
  publisher = {American Physical Society},
  doi = {10.1103/PhysRevB.63.144433},
  url = {https://link.aps.org/doi/10.1103/PhysRevB.63.144433}
}

@article{FurukawaPhysRevB.86.094417,
  title = {Ground-state phase diagram of a spin-$\frac{1}{2}$ frustrated ferromagnetic {XXZ} chain: Haldane dimer phase and gapped/gapless chiral phases},
  author = {Furukawa, Shunsuke and Sato, Masahiro and Onoda, Shigeki and Furusaki, Akira},
  journal = {Phys. Rev. B},
  volume = {86},
  issue = {9},
  pages = {094417},
  numpages = {19},
  year = {2012},
  month = {Sep},
  publisher = {American Physical Society},
  doi = {10.1103/PhysRevB.86.094417},
  url = {https://link.aps.org/doi/10.1103/PhysRevB.86.094417}
}

@article{ChitraPhysRevB.52.6581,
  title = {Density-matrix renormalization-group studies of the spin-1/2 {H}eisenberg system with dimerization and frustration},
  author = {Chitra, R. and Pati, Swapan and Krishnamurthy, H. R. and Sen, Diptiman and Ramasesha, S.},
  journal = {Phys. Rev. B},
  volume = {52},
  issue = {9},
  pages = {6581--6587},
  numpages = {0},
  year = {1995},
  month = {Sep},
  publisher = {American Physical Society},
  doi = {10.1103/PhysRevB.52.6581},
  url = {https://link.aps.org/doi/10.1103/PhysRevB.52.6581}
}

@article{BeraNatComm2022,
title = {Emergent many-body composite excitations of interacting spin-1/2 trimers},
journal = {Nature Communications},
volume = {13},
pages = {6888},
year = {2022},
issn = {2041-1723},
doi = {10.1038/s41467-022-34342-1},
url = {https://doi.org/10.1038/s41467-022-34342-1},
author = {Bera, Anup Kumar and Yusuf, S. M. and Saha, Sudip Kumar and Kumar, Manoranjan and Voneshen, David and Skourski, Yurii and Zvyagin, Sergei A.}
}

@article{MaPhysRevX.13.031016,
  title = {Average {S}ymmetry-{P}rotected {T}opological {P}hases},
  author = {Ma, Ruochen and Wang, Chong},
  journal = {Phys. Rev. X},
  volume = {13},
  issue = {3},
  pages = {031016},
  numpages = {24},
  year = {2023},
  month = {Aug},
  publisher = {American Physical Society},
  doi = {10.1103/PhysRevX.13.031016},
  url = {https://link.aps.org/doi/10.1103/PhysRevX.13.031016}
}

@article{Songsciadv.aao4748,
title = {Observation of symmetry-protected topological band with ultracold fermions},
author = {Bo Song  and Long Zhang  and Chengdong He  and Ting Fung Jeffrey Poon  and Elnur Hajiyev  and Shanchao Zhang  and Xiong-Jun Liu  and Gyu-Boong Jo },
journal = {Science Advances},
volume = {4},
number = {2},
pages = {eaao4748},
year = {2018},
doi = {10.1126/sciadv.aao4748},
URL = {https://www.science.org/doi/abs/10.1126/sciadv.aao4748}
}

@article{LiPhysRevLett.127.263004,
  title = {{S}ymmetry-{P}rotected {T}opological {P}hases in a {R}ydberg {G}lass},
  author = {Li, Kai and Wang, Jiong-Hao and Yang, Yan-Bin and Xu, Yong},
  journal = {Phys. Rev. Lett.},
  volume = {127},
  issue = {26},
  pages = {263004},
  numpages = {6},
  year = {2021},
  month = {Dec},
  publisher = {American Physical Society},
  doi = {10.1103/PhysRevLett.127.263004},
  url = {https://link.aps.org/doi/10.1103/PhysRevLett.127.263004}
}

@article{PollmannPhysRevB.86.125441,
  title = {Detection of symmetry-protected topological phases in one dimension},
  author = {Pollmann, Frank and Turner, Ari M.},
  journal = {Phys. Rev. B},
  volume = {86},
  issue = {12},
  pages = {125441},
  numpages = {13},
  year = {2012},
  month = {Sep},
  publisher = {American Physical Society},
  doi = {10.1103/PhysRevB.86.125441},
  url = {https://link.aps.org/doi/10.1103/PhysRevB.86.125441}
}

@article{HidaPhysRevB.45.2207,
  title = {Crossover between the {H}aldane-gap phase and the dimer phase in the spin-1/2 alternating {H}eisenberg chain},
  author = {Hida, Kazuo},
  journal = {Phys. Rev. B},
  volume = {45},
  issue = {5},
  pages = {2207--2212},
  numpages = {0},
  year = {1992},
  month = {Feb},
  publisher = {American Physical Society},
  doi = {10.1103/PhysRevB.45.2207},
  url = {https://link.aps.org/doi/10.1103/PhysRevB.45.2207}
}

@article{WangPhysRevB.105.184429,
  title = {Hydrodynamics of interacting spinons in the magnetized spin-$\frac{1}{2}$ chain with a uniform {D}zyaloshinskii-{M}oriya interaction},
  author = {Wang, Ren-Bo and Keselman, Anna and Starykh, Oleg A.},
  journal = {Phys. Rev. B},
  volume = {105},
  issue = {18},
  pages = {184429},
  numpages = {16},
  year = {2022},
  month = {May},
  publisher = {American Physical Society},
  doi = {10.1103/PhysRevB.105.184429},
  url = {https://link.aps.org/doi/10.1103/PhysRevB.105.184429}
}

@article{BalentsPhysRevB.101.020401,
  title = {Collective spinon spin wave in a magnetized {U}(1) spin liquid},
  author = {Balents, Leon and Starykh, Oleg A.},
  journal = {Phys. Rev. B},
  volume = {101},
  issue = {2},
  pages = {020401},
  numpages = {5},
  year = {2020},
  month = {Jan},
  publisher = {American Physical Society},
  doi = {10.1103/PhysRevB.101.020401},
  url = {https://link.aps.org/doi/10.1103/PhysRevB.101.020401}
}

@article{RejaPhysRevB.99.134420,
  title = {Critical spin-$\frac{1}{2}$ tetramer compound {C}u{I}n{VO}$_{5}$: Exploring the vicinity of two multimerized singlet states},
  author = {Reja, Sahinur and Nishimoto, Satoshi},
  journal = {Phys. Rev. B},
  volume = {99},
  issue = {13},
  pages = {134420},
  numpages = {9},
  year = {2019},
  month = {Apr},
  publisher = {American Physical Society},
  doi = {10.1103/PhysRevB.99.134420},
  url = {https://link.aps.org/doi/10.1103/PhysRevB.99.134420}
}

@article{KramersPhysRev.60.252,
  title = {Statistics of the {T}wo-{D}imensional {F}erromagnet. {P}art {I}},
  author = {Kramers, H. A. and Wannier, G. H.},
  journal = {Phys. Rev.},
  volume = {60},
  issue = {3},
  pages = {252},
  numpages = {0},
  year = {1941},
  month = {Aug},
  publisher = {American Physical Society},
  doi = {10.1103/PhysRev.60.252},
  url = {https://link.aps.org/doi/10.1103/PhysRev.60.252}
}

@book{giamarchi,
  author = {Giamarchi, Thierry},
  title = {Quantum Physics in One Dimension},
  publisher = {Oxford University Press},
  year = {2003},
  month = {12},
  isbn = {9780198525004},
  doi = {10.1093/acprof:oso/9780198525004.001.0001},
  url = {https://doi.org/10.1093/acprof:oso/9780198525004.001.0001}
}

@article{Léséleucscience.aav9105,
author = {Sylvain de Léséleuc and Vincent Lienhard  and Pascal Scholl  and Daniel Barredo  and Sebastian Weber  and Nicolai Lang  and Hans Peter Büchler  and Thierry Lahaye  and Antoine Browaeys },
title = {Observation of a symmetry-protected topological phase of interacting bosons with {R}ydberg atoms},
journal = {Science},
volume = {365},
number = {6455},
pages = {775-780},
year = {2019},
doi = {10.1126/science.aav9105},
URL = {https://www.science.org/doi/abs/10.1126/science.aav9105}
}

@article{HasePhysRevB.94.174421,
  title = {Magnetism of the antiferromagnetic spin-$\frac{1}{2}$ tetramer compound ${\mathrm{{C}u{I}n{VO}}}_{5}$},
  author = {Hase, Masashi and Matsumoto, Masashige and Matsuo, Akira and Kindo, Koichi},
  journal = {Phys. Rev. B},
  volume = {94},
  issue = {17},
  pages = {174421},
  numpages = {7},
  year = {2016},
  month = {Nov},
  publisher = {American Physical Society},
  doi = {10.1103/PhysRevB.94.174421},
  url = {https://link.aps.org/doi/10.1103/PhysRevB.94.174421}
}

@article{KargarianPhysRevA.77.032346,
  title = {Renormalization of entanglement in the anisotropic {H}eisenberg $({XXZ})$ model},
  author = {Kargarian, M. and Jafari, R. and Langari, A.},
  journal = {Phys. Rev. A},
  volume = {77},
  issue = {3},
  pages = {032346},
  numpages = {8},
  year = {2008},
  month = {Mar},
  publisher = {American Physical Society},
  doi = {10.1103/PhysRevA.77.032346},
  url = {https://link.aps.org/doi/10.1103/PhysRevA.77.032346}
}

@article{SchlappaNatComm2018,
title = {Probing multi-spinon excitations outside of the two-spinon continuum in the antiferromagnetic spin chain cuprate $\mathrm{{S}r}_2\mathrm{{C}u}\mathrm{{O}}_3$},
journal = {Nature Communications},
volume = {9},
pages = {5394},
year = {2018},
issn = {2041-1723},
doi = {10.1038/s41467-018-07838-y},
url = {https://doi.org/10.1038/s41467-018-07838-y},
author = {Schlappa, J. and Kumar, U. and Zhou, K. J. and Singh, S. and Mourigal, M. and Strocov, V. N. and Revcolevschi, A. and Patthey, L. and Rønnow, H. M. and Johnston, S. and Schmitt, T.}
}

@article{KumarPhysRevB.102.075134,
  title = {Spectroscopic signatures of next-nearest-neighbor hopping in the charge and spin dynamics of doped one-dimensional antiferromagnets},
  author = {Kumar, Umesh and Nocera, Alberto and Price, Gregory and Stiwinter, Kenneth and Johnston, Steven and Datta, Trinanjan},
  journal = {Phys. Rev. B},
  volume = {102},
  issue = {7},
  pages = {075134},
  numpages = {12},
  year = {2020},
  month = {Aug},
  publisher = {American Physical Society},
  doi = {10.1103/PhysRevB.102.075134},
  url = {https://link.aps.org/doi/10.1103/PhysRevB.102.075134}
}

@article{NagNatComm2022,
  title = {Quadrupolar magnetic excitations in an isotropic spin-1 antiferromagnet},
  author = {Nag, A and Nocera, A. and Agrestini, S. and Garcia-Fernandez, M. and Walters, A. C. and Cheong, Sang-Wook and Johnston, S. and Zhou, Ke-Jin},
  journal = {Nature Communications},
  volume = {13},
  issue = {1},
  pages = {2327},
  year = {2022},
  doi = {10.1038/s41467-022-30065-5},
  url = {https://doi.org/10.1038/s41467-022-30065-5}
}

@article{NoceraSciRep2018,
  title = {{C}omputing {R}esonant {I}nelastic {X}-{R}ay {S}cattering {S}pectra {U}sing {T}he {D}ensity {M}atrix {R}enormalization {G}roup {M}ethod},
  author = {Nocera, A. and Kumar, U. and Kaushal, N. and Alvarez, G. and Dagotto, E. and Johnston, S.},
  journal = {Scientific Reports},
  volume = {8},
  issue = {1},
  pages = {11080},
  year = {2018},
  doi = {10.1038/s41598-018-29218-8},
  url = {https://doi.org/10.1038/s41598-018-29218-8}
}

@article{GaoPhysRevB.109.L020402,
  title = {Spinon continuum in the {H}eisenberg quantum chain compound {S}r$_{2}${V}$_{3}${O}$_{9}$},
  author = {Gao, Shang and Lin, Ling-Fang and Laurell, Pontus and Chen, Qiang and Huang, Qing and Cruz, Clarina dela and Vemuru, Krishnamurthy V. and Lumsden, Mark D. and Nagler, Stephen E. and Alvarez, Gonzalo and Dagotto, Elbio and Zhou, Haidong and Christianson, Andrew D. and Stone, Matthew B.},
  journal = {Phys. Rev. B},
  volume = {109},
  issue = {2},
  pages = {L020402},
  numpages = {8},
  year = {2024},
  month = {Jan},
  publisher = {American Physical Society},
  doi = {10.1103/PhysRevB.109.L020402},
  url = {https://link.aps.org/doi/10.1103/PhysRevB.109.L020402}
}

@article{NoceraPhysRevE.94.053308,
  title = {Spectral functions with the density matrix renormalization group: Krylov-space approach for correction vectors},
  author = {A. Nocera and G. Alvarez},
  journal = {Phys. Rev. E},
  volume = {94},
  issue = {5},
  pages = {053308},
  numpages = {8},
  year = {2016},
  month = {Nov},
  publisher = {American Physical Society},
  doi = {10.1103/PhysRevE.94.053308},
  url = {https://link.aps.org/doi/10.1103/PhysRevE.94.053308}
}

@article{Schmiedinghoff2022,
      title={Three-body bound states in antiferromagnetic spin ladders}, 
      journal = {Commun. Phys.},
      volume = {5},
      pages = {218},
      year={2022},
      issn = {2399-3650},
      doi = {10.1038/s42005-022-00986-0},
      url={https://doi.org/10.1038/s42005-022-00986-0},
      author={Schmiedinghoff, Gary and Müller, Leanna and Kumar, Umesh and Uhrig, Götz S. and Fauseweh, Benedikt}
}

@article{SinghaniaPhysRevB.98.104429,
  title = {Cluster mean-field study of the {H}eisenberg model for {C}u{I}n{VO}$_{5}$},
  author = {Singhania, Ayushi and Kumar, Sanjeev},
  journal = {Phys. Rev. B},
  volume = {98},
  issue = {10},
  pages = {104429},
  numpages = {9},
  year = {2018},
  month = {Sep},
  publisher = {American Physical Society},
  doi = {10.1103/PhysRevB.98.104429},
  url = {https://link.aps.org/doi/10.1103/PhysRevB.98.104429}
}

@article{PrabhakaPhysRevB.111.205106,
  title = {Fractionalized excitations and resonant inelastic x-ray spectra in frustrated spin-1/2 trimer chains},
  author = {Prabhakar and Pal, Subhajyoti and Kumar, Umesh and Kumar, Manoranjan and Mukherjee, Anamitra},
  journal = {Phys. Rev. B},
  volume = {111},
  issue = {20},
  pages = {205106},
  numpages = {12},
  year = {2025},
  month = {May},
  publisher = {American Physical Society},
  doi = {10.1103/PhysRevB.111.205106},
  url = {https://link.aps.org/doi/10.1103/PhysRevB.111.205106}
}

@article{JafariPhysRevB.76.014412,
  title = {Phase diagram of the one-dimensional ${S}=\frac{1}{2}$ ${XXZ}$ model with ferromagnetic nearest-neighbor and antiferromagnetic next-nearest-neighbor interactions},
  author = {Jafari, R. and Langari, A.},
  journal = {Phys. Rev. B},
  volume = {76},
  issue = {1},
  pages = {014412},
  numpages = {7},
  year = {2007},
  month = {Jul},
  publisher = {American Physical Society},
  doi = {10.1103/PhysRevB.76.014412},
  url = {https://link.aps.org/doi/10.1103/PhysRevB.76.014412}
}

@article{KuhnerPhysRevB.60.335,
  title = {Dynamical correlation functions using the density matrix renormalization group},
  author = {K\"uhner, Till D. and White, Steven R.},
  journal = {Phys. Rev. B},
  volume = {60},
  issue = {1},
  pages = {335--343},
  numpages = {0},
  year = {1999},
  month = {Jul},
  publisher = {American Physical Society},
  doi = {10.1103/PhysRevB.60.335},
  url = {https://link.aps.org/doi/10.1103/PhysRevB.60.335}
}

@article{JeckelmannPhysRevB.66.045114,
  title = {Dynamical density-matrix renormalization-group method},
  author = {Jeckelmann, Eric},
  journal = {Phys. Rev. B},
  volume = {66},
  issue = {4},
  pages = {045114},
  numpages = {16},
  year = {2002},
  month = {Jul},
  publisher = {American Physical Society},
  doi = {10.1103/PhysRevB.66.045114},
  url = {https://link.aps.org/doi/10.1103/PhysRevB.66.045114}
}

@article{UmmethumPhysRevB.86.104403,
  title = {Discrete antiferromagnetic spin-wave excitations in the giant ferric wheel {F}e$_{18}$},
  author = {Ummethum, J. and Nehrkorn, J. and Mukherjee, S. and Ivanov, N. B. and Stuiber, S. and Str\"assle, Th. and Tregenna-Piggott, P. L. W. and Mutka, H. and Christou, G. and Waldmann, O. and Schnack, J.},
  journal = {Phys. Rev. B},
  volume = {86},
  issue = {10},
  pages = {104403},
  numpages = {14},
  year = {2012},
  month = {Sep},
  publisher = {American Physical Society},
  doi = {10.1103/PhysRevB.86.104403},
  url = {https://link.aps.org/doi/10.1103/PhysRevB.86.104403}
}

@article{KatsuraPhysRevLett.104.066403,
  title = {Theory of the Thermal Hall Effect in Quantum Magnets},
  author = {Katsura, Hosho and Nagaosa, Naoto and Lee, Patrick A.},
  journal = {Phys. Rev. Lett.},
  volume = {104},
  issue = {6},
  pages = {066403},
  numpages = {4},
  year = {2010},
  month = {Feb},
  publisher = {American Physical Society},
  doi = {10.1103/PhysRevLett.104.066403},
  url = {https://link.aps.org/doi/10.1103/PhysRevLett.104.066403}
}

@article{ChenPhysRevX.8.041028,
  title = {Topological Spin Excitations in Honeycomb Ferromagnet {C}r{I}$_{3}$},
  author = {Chen, Lebing and Chung, Jae-Ho and Gao, Bin and Chen, Tong and Stone, Matthew B. and Kolesnikov, Alexander I. and Huang, Qingzhen and Dai, Pengcheng},
  journal = {Phys. Rev. X},
  volume = {8},
  issue = {4},
  pages = {041028},
  numpages = {7},
  year = {2018},
  month = {Nov},
  publisher = {American Physical Society},
  doi = {10.1103/PhysRevX.8.041028},
  url = {https://link.aps.org/doi/10.1103/PhysRevX.8.041028}
}

@article{Owerre_2016,
doi = {10.1088/0953-8984/28/38/386001},
url = {https://dx.doi.org/10.1088/0953-8984/28/38/386001},
year = {2016},
month = {jul},
publisher = {IOP Publishing},
volume = {28},
number = {38},
pages = {386001},
author = {Owerre, S A},
title = {A first theoretical realization of honeycomb topological magnon insulator},
journal = {Journal of Physics: Condensed Matter}
}

@article{LeePhysRevB.97.180401,
  title = {Magnonic quantum spin Hall state in the zigzag and stripe phases of the antiferromagnetic honeycomb lattice},
  author = {Lee, Ki Hoon and Chung, Suk Bum and Park, Kisoo and Park, Je-Geun},
  journal = {Phys. Rev. B},
  volume = {97},
  issue = {18},
  pages = {180401},
  numpages = {5},
  year = {2018},
  month = {May},
  publisher = {American Physical Society},
  doi = {10.1103/PhysRevB.97.180401},
  url = {https://link.aps.org/doi/10.1103/PhysRevB.97.180401}
}

@article{OwerrePhysRevB.95.014422,
  title = {Topological thermal Hall effect in frustrated kagome antiferromagnets},
  author = {Owerre, S. A.},
  journal = {Phys. Rev. B},
  volume = {95},
  issue = {1},
  pages = {014422},
  numpages = {6},
  year = {2017},
  month = {Jan},
  publisher = {American Physical Society},
  doi = {10.1103/PhysRevB.95.014422},
  url = {https://link.aps.org/doi/10.1103/PhysRevB.95.014422}
}

@article{ZhangPhysRevB.103.174402,
  title = {Topological magnons for thermal Hall transport in frustrated magnets with bond-dependent interactions},
  author = {Zhang, Emily Z. and Chern, Li Ern and Kim, Yong Baek},
  journal = {Phys. Rev. B},
  volume = {103},
  issue = {17},
  pages = {174402},
  numpages = {9},
  year = {2021},
  month = {May},
  publisher = {American Physical Society},
  doi = {10.1103/PhysRevB.103.174402},
  url = {https://link.aps.org/doi/10.1103/PhysRevB.103.174402}
}

@article{SheikhiPhysRevB.104.045139,
  title = {Hybrid topological magnon-phonon modes in ferromagnetic honeycomb and kagome lattices},
  author = {Sheikhi, Bahman and Kargarian, Mehdi and Langari, Abdollah},
  journal = {Phys. Rev. B},
  volume = {104},
  issue = {4},
  pages = {045139},
  numpages = {15},
  year = {2021},
  month = {Jul},
  publisher = {American Physical Society},
  doi = {10.1103/PhysRevB.104.045139},
  url = {https://link.aps.org/doi/10.1103/PhysRevB.104.045139}
}

@article{HungPhysRevLett.114.076401,
  title = {Ground-State Degeneracy of Topological Phases on Open Surfaces},
  author = {Hung, Ling-Yan and Wan, Yidun},
  journal = {Phys. Rev. Lett.},
  volume = {114},
  issue = {7},
  pages = {076401},
  numpages = {5},
  year = {2015},
  month = {Feb},
  publisher = {American Physical Society},
  doi = {10.1103/PhysRevLett.114.076401},
  url = {https://link.aps.org/doi/10.1103/PhysRevLett.114.076401}
}

@article{LuPhysRevLett.125.116801,
  title = {Detecting Topological Order at Finite Temperature Using Entanglement Negativity},
  author = {Lu, Tsung-Cheng and Hsieh, Timothy H. and Grover, Tarun},
  journal = {Phys. Rev. Lett.},
  volume = {125},
  issue = {11},
  pages = {116801},
  numpages = {6},
  year = {2020},
  month = {Sep},
  publisher = {American Physical Society},
  doi = {10.1103/PhysRevLett.125.116801},
  url = {https://link.aps.org/doi/10.1103/PhysRevLett.125.116801}
}

@article{WangPRXQuantum.6.010314,
  title = {Intrinsic Mixed-State Topological Order},
  author = {Wang, Zijian and Wu, Zhengzhi and Wang, Zhong},
  journal = {PRX Quantum},
  volume = {6},
  issue = {1},
  pages = {010314},
  numpages = {29},
  year = {2025},
  month = {Jan},
  publisher = {American Physical Society},
  doi = {10.1103/PRXQuantum.6.010314},
  url = {https://link.aps.org/doi/10.1103/PRXQuantum.6.010314}
}

@article{WhitePhysRevB.77.134437,
  title = {Spectral function for the ${S}=1$ {H}eisenberg antiferromagetic chain},
  author = {S. R. White and I. Affleck},
  journal = {Phys. Rev. B},
  volume = {77},
  issue = {13},
  pages = {134437},
  numpages = {11},
  year = {2008},
  month = {Apr},
  publisher = {American Physical Society},
  doi = {10.1103/PhysRevB.77.134437},
  url = {https://link.aps.org/doi/10.1103/PhysRevB.77.134437}
}

@article{SharmaPhysRevB.111.064404,
  title = {Bound states and deconfined spinons in the dynamical structure factor of the ${J}_{1}\ensuremath{-}{J}_{2}$ spin-1 chain},
  author = {A. Sharma and M. Nayak and H. M. R\o{}nnow and F. Mila},
  journal = {Phys. Rev. B},
  volume = {111},
  issue = {6},
  pages = {064404},
  numpages = {15},
  year = {2025},
  month = {Feb},
  publisher = {American Physical Society},
  doi = {10.1103/PhysRevB.111.064404},
  url = {https://link.aps.org/doi/10.1103/PhysRevB.111.064404}
}

@article{Brink_2007,
doi = {10.1209/0295-5075/80/47003},
url = {https://dx.doi.org/10.1209/0295-5075/80/47003},
year = {2007},
month = {oct},
publisher = {},
volume = {80},
number = {4},
pages = {47003},
author = {J. van den Brink},
title = {The theory of indirect resonant inelastic {X}-ray scattering on magnons},
journal = {Europhysics Letters}
}

@article{FortePhysRevB.83.245133,
  title = {Doping dependence of magnetic excitations of one-dimensional cuprates as probed by resonant inelastic x-ray scattering},
  author = {F. Forte and M. Cuoco and C. Noce and J. van den Brink},
  journal = {Phys. Rev. B},
  volume = {83},
  issue = {24},
  pages = {245133},
  numpages = {7},
  year = {2011},
  month = {Jun},
  publisher = {American Physical Society},
  doi = {10.1103/PhysRevB.83.245133},
  url = {https://link.aps.org/doi/10.1103/PhysRevB.83.245133}
}

@article{PollmannPhysRevB.81.064439,
  title = {Entanglement spectrum of a topological phase in one dimension},
  author = {F. Pollmann and A. M. Turner and E. Berg and M. Oshikawa},
  journal = {Phys. Rev. B},
  volume = {81},
  issue = {6},
  pages = {064439},
  numpages = {10},
  year = {2010},
  month = {Feb},
  publisher = {American Physical Society},
  doi = {10.1103/PhysRevB.81.064439},
  url = {https://link.aps.org/doi/10.1103/PhysRevB.81.064439}
}

@article{Senthilscience.1091806,
author = {T. Senthil and A. Vishwanath and L. Balents and S. Sachdev and M. P. A. Fisher},
title = {Deconfined Quantum Critical Points},
journal = {Science},
volume = {303},
number = {5663},
pages = {1490-1494},
year = {2004},
doi = {10.1126/science.1091806},
URL = {https://www.science.org/doi/abs/10.1126/science.1091806}
}

@article{JiangPhysRevB.99.075103,
  title = {Ising ferromagnet to valence bond solid transition in a one-dimensional spin chain: Analogies to deconfined quantum critical points},
  author = {S. Jiang and O. Motrunich},
  journal = {Phys. Rev. B},
  volume = {99},
  issue = {7},
  pages = {075103},
  numpages = {31},
  year = {2019},
  month = {Feb},
  publisher = {American Physical Society},
  doi = {10.1103/PhysRevB.99.075103},
  url = {https://link.aps.org/doi/10.1103/PhysRevB.99.075103}
}

@article{RobertsPhysRevB.99.165143,
  title = {Deconfined quantum critical point in one dimension},
  author = {B. Roberts and S. Jiang and O. I. Motrunich},
  journal = {Phys. Rev. B},
  volume = {99},
  issue = {16},
  pages = {165143},
  numpages = {19},
  year = {2019},
  month = {Apr},
  publisher = {American Physical Society},
  doi = {10.1103/PhysRevB.99.165143},
  url = {https://link.aps.org/doi/10.1103/PhysRevB.99.165143}
}

@article{HuangPhysRevB.100.125137,
  title = {Emergent symmetry and conserved current at a one-dimensional incarnation of deconfined quantum critical point},
  author = {R.-Z. Huang and D.-C. Lu and Y.-Z. You and Z. Y. Meng and T. Xiang},
  journal = {Phys. Rev. B},
  volume = {100},
  issue = {12},
  pages = {125137},
  numpages = {16},
  year = {2019},
  month = {Sep},
  publisher = {American Physical Society},
  doi = {10.1103/PhysRevB.100.125137},
  url = {https://link.aps.org/doi/10.1103/PhysRevB.100.125137}
}

@article{YangPhysRevE.104.064121,
  title = {Quantum entanglement and criticality in a one-dimensional deconfined quantum critical point},
  author = {S. Yang and J.-B. Xu},
  journal = {Phys. Rev. E},
  volume = {104},
  issue = {6},
  pages = {064121},
  numpages = {10},
  year = {2021},
  month = {Dec},
  publisher = {American Physical Society},
  doi = {10.1103/PhysRevE.104.064121},
  url = {https://link.aps.org/doi/10.1103/PhysRevE.104.064121}
}

@article{LiPhysRevB.107.085130,
  title = {Exploring dynamical quantum phase transitions in a spin model with deconfined critical point via the quantum steering ellipsoid},
  author = {C.-X. Li and S. Yang and J.-B. Xu and H.-Q. Lin},
  journal = {Phys. Rev. B},
  volume = {107},
  issue = {8},
  pages = {085130},
  numpages = {10},
  year = {2023},
  month = {Feb},
  publisher = {American Physical Society},
  doi = {10.1103/PhysRevB.107.085130},
  url = {https://link.aps.org/doi/10.1103/PhysRevB.107.085130}
}

@article{Pronks53y-qmr4,
  title = {Deconfined quantum criticality in a frustrated {H}aldane chain with single-ion anisotropy},
  author = {N. T. Pronk and B. M. La Rivi\`ere and N. Chepiga},
  journal = {Phys. Rev. B},
  volume = {111},
  issue = {22},
  pages = {L220412},
  numpages = {7},
  year = {2025},
  month = {Jun},
  publisher = {American Physical Society},
  doi = {10.1103/s53y-qmr4},
  url = {https://link.aps.org/doi/10.1103/s53y-qmr4}
}

@article{FortePhysRevB.77.134428,
  title = {Magnetic excitations in {L}a$_{2}${C}u{O}$_{4}$ probed by indirect resonant inelastic x-ray scattering},
  author = {Forte, Filomena and Ament, Luuk J. P. and van den Brink, Jeroen},
  journal = {Phys. Rev. B},
  volume = {77},
  issue = {13},
  pages = {134428},
  numpages = {9},
  year = {2008},
  month = {Apr},
  publisher = {American Physical Society},
  doi = {10.1103/PhysRevB.77.134428},
  url = {https://link.aps.org/doi/10.1103/PhysRevB.77.134428}
}

@article{IshiiPhysRevB.112.195132,
  title = {Observation of spin-conserving two-spinon continuum in the ${S}=\frac{1}{2}$ antiferromagnetic chain system {S}r$_{2}${C}u{O}$_{3}$ using {C}u ${K}$-edge resonant inelastic x-ray scattering},
  author = {Ishii, Kenji and Tsutsui, Kenji and Kawamata, Takayuki and Koike, Yoji},
  journal = {Phys. Rev. B},
  volume = {112},
  issue = {19},
  pages = {195132},
  numpages = {9},
  year = {2025},
  month = {Nov},
  publisher = {American Physical Society},
  doi = {10.1103/53wr-vpcp},
  url = {https://link.aps.org/doi/10.1103/53wr-vpcp}
}


\end{document}